\documentclass[]{mn2e}
\usepackage{epsfig}
\usepackage{graphicx}
\usepackage{tikz}
\usepackage{pgf}

\title[M\,33 monitoring. IV]{The UK Infrared Telescope M\,33 monitoring
project. IV. Variable red giant stars across the galactic disc}
\author[Javadi et al.]{\parbox{17cm}{
                       Atefeh Javadi$^{1}$,
                       Maryam Saberi$^{3}$, 
                       Jacco Th.\ van Loon$^{2}$,
                       Habib Khosroshahi$^{1}$,
                       Najmeh Golabatooni$^{3}$,
                        and
                       Mohammad Taghi Mirtorabi$^{3}$}\\
$^{1}$School of Astronomy, Institute for Research in Fundamental Sciences
      (IPM), P.O.\ Box 19395-5531, Tehran, Iran\\
$^{2}$Astrophysics Group, Lennard-Jones Laboratories, Keele University,
      Staffordshire ST5 5BG, UK\\
$^{3}$Physics Department, Alzahra University, Vanak, 1993891176, Tehran, Iran}
\date{Resubmitted: 2014}
\pagerange{\pageref{firstpage}--\pageref{lastpage}}
\pubyear{2014}
\begin{document}
\maketitle
\label{firstpage}
\begin{abstract}
We have conducted a near-infrared monitoring campaign at the UK InfraRed 
Telescope (UKIRT), of the Local Group spiral galaxy M\,33 (Triangulum). The
main aim was to identify stars in the very final stage of their evolution, and
for which the luminosity is more directly related to the birth mass than the
more numerous less-evolved giant stars that continue to increase in
luminosity. In this fourth paper of the series, we present a search for
variable red giant stars in an almost square degree region comprising most of
the galaxy's disc, carried out with the WFCAM instrument in the K band. These
data, taken during the period 2005--2007, were complemented by J- and H-band
images. Photometry was obtained for 403\,734 stars in this region; of these,
4643 stars were found to be variable, most of which are Asymptotic Giant
Branch (AGB) stars. The variable stars are concentrated towards the centre of
M\,33, more so than low-mass, less-evolved red giants. Our data were matched
to optical catalogues of variable stars and carbon stars and to mid-infrared
photometry from the {\it Spitzer} Space Telescope. Most dusty AGB stars had
not been previously identified in optical variability surveys, and our survey
is also more complete for these types of stars than the {\it Spitzer} survey.
The photometric catalogue is made publicly available at the Centre de
Donn\'ees astronomiques de Strasbourg.
\end{abstract}

\section{Introduction}

The Local Group galaxy Triangulum (Hodierna 1654) -- hereafter referred to as
M\,33 (Messier 1771) -- offers us a unique opportunity to study stellar
populations, their history and their feedback across an entire spiral galaxy
and in particular in its central regions, that in our own Milky Way are
heavily obscured by the intervening dusty disc (van Loon et al.\ 2003;
Benjamin et al.\ 2005). Our viewing angle with respect to the M\,33 disc is
more favourable (56--$57^\circ$ -- Zaritsky, Elston \& Hill 1989; Deul \& van
der Hulst 1987) than that of the larger M\,31 (Andromeda), whilst the distance
to M\,33 is not much different from that to M\,31 ($\mu=24.9$ mag -- Bonanos
et al.\ 2006). For these reasons, numerous surveys have been conducted to
study M\,33 in different wavelength ranges including optical (Macri et al.\
2001; Hartman et al.\ 2006; Massey et al.\ 2006); radio (Engargiola et al.\
2003; Rosolowski et al.\ 2007); X-ray (Pietsch et al.\ 2004); and infrared
(IR -- Two Micron All Sky Survey [2MASS], Skrutskie et al.\ 2006; McQuinn et
al.\ 2007). Large populations of Asymptotic Giant Branch (AGB) stars have been
identified in M\,33 (Cioni et al.\ 2008), as well as red supergiants (RSGs) up
to progenitor masses in excess of 20 M$_\odot$ (Drout, Massey \& Meynet 2012).
Many of them are dusty Long Period Variables (LPVs -- McQuinn et al.\ 2007;
Thompson et al.\ 2009), and these have been found also in the central parts of
M\,33 (Javadi, van Loon \& Mirtorabi 2011a; Javadi et al.\ 2013).

In the final stage of stellar evolution, low--medium mass (0.8--8 M$_\odot$)
stars enter the AGB phase (Marigo et al.\ 2008) and more massive stars ($M>8$
M$_\odot$) enter the RSG phase (Levesque et al.\ 2005; Levesque 2010). These
two phases of evolution trace stellar populations over a range in age from
$\sim10$ Myr to $\sim10$ Gyr, and hence the evolution of their host galaxies
over essential all cosmological time. Radial pulsations of the atmospheric
layers in AGB stars and RSGs yield long-period variability of order 150--1500
days in the photometric light curves (e.g., Wood et al.\ 1992; Wood 1998;
Pierce et al.\ 2000; Whitelock et al.\ 2003). Most evolved AGB stars pulsate
in the fundamental mode, while less evolved AGB stars and RSGs pulsate in an
over-tone; as a result the amplitude of variability expressed in magnitude of
AGB stars are larger than that of RSGs and less evolved AGB stars. These LPVs
are powerful tools to study the star formation history of galaxies, and to
this aim various variability surveys have been conducted of M\,33 over recent
years (Macri et al.\ 2001; Mochejska et al.\ 2001a,b; Hartman et al.\ 2006;
Sarajedini et al.\ 2006; McQuinn et al.\ 2007).

The coolest ($T\sim3000$--4000 K) and most luminous ($\sim10\,000$--60\,000
L$_\odot$) AGB stars create large amounts of dust in the star’s atmosphere.
This dust is ejected into space and can cloak the star, especially in the
optical light, and add luminosity at IR wavelengths. Likewise, RSGs stand out
especially at IR wavelengths. Hence, among all surveys, those with the {\it
Spitzer} Space Telescope (McQuinn et al.\ 2007) and the UK InfraRed Telescope
(UKIRT, Javadi et al.\ 2011a) were more successful in detecting dusty LPVs.
Since these evolved stars shed a large amount of mass into the interstellar
medium (ISM), they are important factors in changing the chemical composition
of galaxies and incrementing the rate of stellar birth.

The main objectives of our project are described in Javadi, van Loon \&
Mirtorabi (2011c): to construct the mass function of LPVs and derive from this
the star formation history in M\,33; to correlate spatial distributions of the
LPVs of different mass with galactic structures (spheroid, disc and spiral arm
components); to measure the rate at which dust is produced and fed into the
ISM; to establish correlations between the dust production rate, luminosity,
and amplitude of an LPV; and to compare the {\it in situ} dust replenishment
with the amount of pre-existing dust. Paper I in the series presented the
photometric catalogue of stars in the inner square kpc (Javadi et al.\ 2011a),
with Paper II presenting the galactic structure and star formation history
(Javadi, van Loon \& Mirtorabi 2011b), and Paper III presenting the mass-loss
mechanism and dust production rate (Javadi et al.\ 2013). This paper describes
the extension of the survey to a nearly square degree area covering much of
the M\,33 optical disc. Subsequent papers in the series will cover the star
formation history and mass return in this enlarged area.

In Section 2 we present the observational data, method of photometry on the
images, and accuracy of these measurements. The methodology and completeness
of our search for LPVs, and characterization of their amplitude of variability
are discussed in Section 3. Section 4 describes the photometric catalogue of
all detected stars, which is made available at the Centre de Donn\'ees
astronomiques de Strasbourg (CDS). In section 5 we describe the properties and
distribution of the detected LPVs and also compare these with other (optical
and IR) variability surveys and stellar catalogues. Section 6 summarizes and
concludes the results.

\section{Observations}

Observations were made with three of UKIRT's imagers: UIST, UFTI and WFCAM.
UIST and UFTI cover the central part ($\approx 1$ kpc$^2$) and the photometry,
star formation history and mass return in this region was explained in Papers
I, II and III. In this sequel we focus on the data from WFCAM, which cover a
much larger part of M\,33.

\subsection{WFCAM}

\begin{table}
\caption[]{Log of our observations of each of four tiles (``Q'').}
\begin{tabular}{cccccc}
\hline\hline
Date (y\,m\,d)            &
Q                         &
Filte\rlap{r}             &
Epoch                     &
$t_{\rm int}$ (min\rlap{)} &
Airmass                   \\
\hline
2005 09 18 & 3 & K & 1 & \llap{2}0.3 & 1.035--1.058 \\
2005 09 18 & 2 & K & 1 & \llap{2}0.3 & 1.072--1.110 \\
2005 09 18 & 4 & K & 1 & \llap{2}0.3 & 1.248--1.338 \\
2005 09 19 & 1 & K & 1 & \llap{2}0.3 & 1.021--1.018 \\
2005 10 18 & 3 & K & 2 & \llap{2}0.3 & 1.019--1.021 \\
2005 10 18 & 2 & K & 2 & \llap{2}0.3 & 1.025--1.040 \\
2005 10 18 & 4 & K & 2 & \llap{2}0.3 & 1.053--1.083 \\
2005 10 18 & 1 & K & 2 & \llap{2}0.3 & 1.101--1.149 \\
2005 11 04 & 1 & K & 3 & \llap{2}0.3 & 1.018--1.023 \\
2005 11 04 & 2 & K & 3 & \llap{1}3.5 & 1.028--1.036 \\
2005 12 23 & 2 & K & 4 & \llap{2}7.0 & 1.019--1.022 \\
2005 12 23 & 3 & K & 3 & \llap{2}0.3 & 1.028--1.046 \\
2006 07 21 & 1 & K & 4 & \llap{2}0.3 & 1.425--1.325 \\
2006 07 21 & 2 & K & 5 & \llap{2}0.3 & 1.287--1.214 \\
2006 07 21 & 3 & K & 4 & \llap{2}0.3 & 1.183--1.132 \\
2006 07 21 & 4 & K & 3 & \llap{2}0.3 & 1.109--1.074 \\
2006 10 28 & 1 & K & 5 & \llap{2}7.0 & 1.294--1.126 \\
2006 10 28 & 1 & J & 1 & \llap{2}0.3 & 1.102--1.076 \\
2006 10 29 & 1 & K & 6 & \llap{2}0.3 & 1.445--1.347 \\
2006 10 29 & 1 & H & 1 & \llap{2}7.0 & 1.295--1.209 \\
2006 10 29 & 1 & J & 2 & \llap{2}7.0 & 1.115--1.062 \\
2006 10 30 & 1 & J & 2 & \llap{3}3.8 & 1.200--1.109 \\
2006 10 30 & 4 & K & 4 & \llap{3}3.8 & 1.085--1.044 \\
2006 10 31 & 4 & H & 1 &         6.8 & 1.301--1.301 \\
2006 12 05 & 2 & K & 6 & \llap{2}0.3 & 1.019--1.025 \\
2006 12 12 & 3 & K & 5 & \llap{2}0.3 & 1.082--1.052 \\
2006 12 12 & 3 & H & 1 & \llap{2}0.3 & 1.040--1.027 \\
2007 01 14 & 1 & K & 7 & \llap{2}0.3 & 1.124--1.169 \\
2007 01 14 & 1 & J & 3 & \llap{2}0.3 & 1.217--1.284 \\
2007 01 14 & 1 & H & 2 & \llap{2}0.3 & 1.342--1.441 \\
2007 01 15 & 2 & K & 7 & \llap{2}0.3 & 1.119--1.163 \\
2007 01 16 & 2 & H & 1 & \llap{2}0.3 & 1.063--1.092 \\
2007 01 17 & 3 & J & 1 & \llap{2}0.3 & 1.031--1.047 \\
2007 01 18 & 2 & J & 2 & \llap{2}0.3 & 1.029--1.044 \\
2007 01 25 & 3 & K & 6 & \llap{2}0.3 & 1.072--1.104 \\
2007 01 25 & 3 & H & 2 & \llap{2}0.3 & 1.161--1.215 \\
2007 09 14 & 1 & K & 8 & \llap{2}0.3 & 1.122--1.086 \\
2007 09 14 & 1 & J & 4 & \llap{1}3.5 & 1.070--1.058 \\
2007 09 14 & 1 & H & 3 & \llap{1}3.5 & 1.046--1.038 \\
2007 09 19 & 2 & K & 8 & \llap{2}0.3 & 1.774--1.606 \\
2007 10 04 & 2 & J & 3 & \llap{1}3.5 & 1.208--1.181 \\
2007 10 04 & 2 & H & 2 & \llap{1}3.5 & 1.155--1.132 \\
2007 10 13 & 3 & K & 7 & \llap{2}0.3 & 1.108--1.076 \\
2007 10 13 & 3 & H & 3 & \llap{1}3.5 & 1.056--1.046 \\
2007 10 24 & 4 & K & 5 & \llap{2}0.3 & 1.135--1.097 \\
2007 10 24 & 3 & J & 2 & \llap{1}3.5 & 1.078--1.064 \\
2007 10 24 & 4 & J & 1 & \llap{1}3.5 & 1.050--1.041 \\
2007 10 24 & 4 & H & 2 & \llap{1}3.5 & 1.025--1.022 \\
\hline
\end{tabular}
\end{table}

The monitoring campaign comprises observations with the Wide Field CAMera
(WFCAM) such that four separately pointed observations (``tiles'', viz.\
M\,33-1, M\,33-2, M\,33-3 and M\,33-4) may be combined to cover a filled
square area of sky covering 0.89 square degree (13 kpc $\times$ 13 kpc) at a
pixel size of $0\rlap{.}^{\prime\prime}4$. The approximate centres of the
camera pointings are respectively ($1^{\rm h}33^{\rm m}19\rlap{.}^{\rm s}30$,
$+30^\circ32^\prime50^{\prime\prime}$), ($1^{\rm h}34^{\rm m}22\rlap{.}^{\rm s}50$,
$+30^\circ32^\prime50^{\prime\prime}$), ($1^{\rm h}34^{\rm m}22\rlap{.}^{\rm s}50$,
$+30^\circ46^\prime23^{\prime\prime}$), ($1^{\rm h}33^{\rm m}19\rlap{.}^{\rm s}30$,
$+30^\circ46^\prime23^{\prime\prime}$). Observations were made in the K band
(UKIRT filter K98) over the period September 2005--October 2007 (Table 1). On
some occasions observations were made also in the J band and/or H band (UKIRT
filters J98 and H98, respectively) to provide colour information. The total
integration on one tile was achieved through four separate exposures. The
number of epochs varies per tile between five and eight, but overlapping
regions will have been observed more frequently.

The large pixel scale of WFCAM means that the point spread function (PSF) is
not always adequately sampled for the crowded fields in M\,33 under modal
observing conditions; to remedy this we employed a $3\times3$ microstepping
scheme to improve the sampling of the PSF. At each position of the nine-point
microstepping sequence a 5-sec exposure was taken, and a small offset
(``jitter'', $\approx0\rlap{.}^{\prime \prime}5$) was applied between each of
nine subsequent microstep sequences. This cycle was repeated three times.
Thus, a typical observation accrued a total integration time of $\approx20$
min; however, in practice fewer or more repeats were performed depending on
conditions, time available, or for technical reasons (interruptions).

The images were processed using the WFCAM pipeline by the Cambridge Astronomy
Survey Unit (CASU -- http://casu.ast.cam.ac.uk/). The main steps include:
\begin{itemize}
\item{Dark current correction;}
\item{Flat field correction to remove pixel sensitivity differences and gain
differences between data channels and between detectors;}
\item{Confidence map generation: a normalized inverse weight map denoting the
confidence associated with the flux values in each pixel;}
\item{Defringing, provided the fringe spatial pattern was available;}
\item{Sky subtraction, either using the ditter sequences themselves (our case)
or using observations of offset sky regions;}
\item{Image persistence and detector crosstalk correction, i.e.\ modelling and
removing electronic effects;}
\item{Combine (``interleave'') the images from the microstepping sequence;}
\item{Shift and average individual images from the jittering sequence;}
\item{Catalogue generation. Objects are identified from the images and listed
in FITS binary tables. This catalogue includes assorted aperture flux
measures, intensity-weighted centroid estimates, and shape information such as
intensity-weighted second moments to encode the equivalent elliptical Gaussian
light distribution;}
\item{Astrometric calibration. All objects in the catalogue are matched to
astrometric standards to define a World Coordinate System (WCS) for each
image/catalogue;}
\item{Photometric zeropoint from comparison of instrumental magnitudes with
2MASS (K band is $K_{\rm s}$).}
\end{itemize}

Our programme IDs are U/05B/18, U/06B/40 and U/07B/17. We have complemented
our data with WFCAM archival data taken for two projects, viz.\ U/05B/7 and
U/05B/H47 which we briefly describe below.

\subsubsection{U/05B/7}

\begin{table}
\caption[]{Log of WFCAM observations of each of four tiles (``Q''), from
programme U/05B/7.}
\begin{tabular}{cccccc}
\hline\hline
Date (y\,m\,d)            &
Q                         &
Filte\rlap{r}             &
Epoch                     &
$t_{\rm int}$ (min\rlap{)} &
Airmass                   \\
\hline
2005 09 29 & 4--1 & J & 1 & 1.0 & 1.196--1.197 \\
2005 09 29 & 4--1 & H & 1 & 4.5 & 1.206--1.227 \\
2005 09 29 & 4--1 & K & 1 & 4.5 & 1.227--1.023 \\
2005 09 30 & 2--1 & J & 1 & 1.0 & 1.023--1.023 \\
2005 09 30 & 2--1 & H & 1 & 4.5 & 1.024--1.024 \\
2005 09 30 & 2--1 & K & 1 & 4.5 & 1.028--1.028 \\
2005 09 30 & 2--3 & J & 1 & 1.0 & 1.055--1.056 \\
2005 09 30 & 2--3 & H & 1 & 4.5 & 1.060--1.060 \\
2005 09 30 & 2--3 & K & 1 & 4.5 & 1.069--1.069 \\
2005 10 24 & 1--1 & J & 1 & 1.0 & 1.335--1.333 \\
2005 10 24 & 1--1 & H & 1 & 4.5 & 1.320--1.319 \\
2005 10 24 & 1--1 & K & 1 & 4.5 & 1.293--1.292 \\
2005 10 24 & 1--2 & J & 1 & 1.0 & 1.050--1.020 \\
2005 10 24 & 1--2 & H & 1 & 4.5 & 1.340--1.210 \\
2005 10 24 & 1--2 & K & 1 & 4.5 & 1.210--1.120 \\
2005 10 24 & 1--3 & J & 1 & 1.0 & 1.204--1.202 \\
2005 10 24 & 1--3 & H & 1 & 4.5 & 1.193--1.193 \\
2005 10 24 & 1--3 & K & 1 & 4.5 & 1.175--1.174 \\
2005 10 24 & 1--4 & J & 1 & 1.0 & 1.158--1.157 \\
2005 10 24 & 1--4 & H & 1 & 4.5 & 1.145--1.144 \\
2005 10 24 & 1--4 & H & 1 & 4.5 & 1.130--1.130 \\
2005 11 05 & 3--1 & K & 1 & 4.5 & 1.073--1.073 \\
2005 11 05 & 3--1 & H & 1 & 4.5 & 1.064--1.064 \\
2005 11 05 & 3--1 & J & 1 & 1.0 & 1.055--1.055 \\
2005 11 05 & 3--2 & K & 1 & 4.5 & 1.052--1.052 \\
2005 11 05 & 3--2 & H & 1 & 4.5 & 1.046--1.046 \\
2005 11 05 & 3--2 & J & 1 & 1.0 & 1.039--1.038 \\
2005 11 27 & 1--1 & J & 2 & 1.0 & 1.292--1.295 \\
2005 12 16 & 1--1 & J & 2 & 1.0 & 1.107--1.106 \\
2005 12 16 & 1--1 & H & 2 & 4.5 & 1.101--1.101 \\
2005 12 16 & 1--1 & K & 2 & 4.5 & 1.090--1.089 \\
2005 12 16 & 1--2 & J & 2 & 1.0 & 1.079--1.078 \\
2005 12 16 & 1--2 & H & 2 & 4.5 & 1.074--1.074 \\
2005 12 16 & 1--2 & K & 2 & 4.5 & 1.065--1.064 \\
2005 12 16 & 1--3 & J & 2 & 1.0 & 1.055--1.054 \\
2005 12 16 & 1--3 & H & 2 & 4.5 & 1.051--1.051 \\
2005 12 16 & 1--3 & K & 2 & 4.5 & 1.044--1.044 \\
2005 12 16 & 1--4 & J & 2 & 1.0 & 1.038--1.038 \\
2005 12 16 & 1--4 & H & 2 & 4.5 & 1.035--1.035 \\
2005 12 16 & 1--4 & K & 2 & 4.5 & 1.031--1.031 \\
\hline
\end{tabular}
\end{table}

The data of this programme (Cioni et al.\ 2008) were taken to survey the
luminous red giant stars of Local Group galaxies. M\,33 was observed on four
occasions: 29 and 30 September, 24 October, 5 November and 16 December 2005
(Table 2). $JHK_{\rm s}$ photometry was obtained from a mosaic of four fields
(instead of the one central field in our case), covering an area $\approx3$
square degrees. The $H$ and $K_{\rm s}$ data were acquired employing a
three-point jitter pattern with $3\times3$ microstepping and 10-sec exposures
per position, giving a total integration time of 270 sec. The $J$ data were
acquired using a five-point jitter pattern with three 10-sec exposures but no
microstepping, resulting in a total integration time of 150 sec. The data were
processed by the CASU.

\subsubsection{U/05B/H47}

\begin{table}
\caption[]{Log of WFCAM observations of each of four tiles (``Q''), from
programme U/05B/H47.}
\begin{tabular}{cccccc}
\hline\hline
Date (y\,m\,d)            &
Q                         &
Filte\rlap{r}             &
Epoch                     &
$t_{\rm int}$ (min\rlap{)} &
Airmass                   \\
\hline
2005 10 20 & 1 & K & 1 & 8.3 & 1.077--1.059 \\
2005 10 20 & 2 & K & 1 & 8.3 & 1.052--1.041 \\
2005 10 20 & 3 & K & 1 & 8.3 & 1.038--1.030 \\
2005 10 20 & 4 & K & 1 & 8.3 & 1.028--1.023 \\
\hline
\end{tabular}
\end{table}

The M\,33 data for this programme (PI: M.\ Irwin) were acquired on 20 October
2005 (Table 3). The covering area is nearly identical to ours, with the camera
pointings called M\,33-position 1, M\,33-position 2, M\,33-position 3 and
M\,33-position 4, respectively.

\subsection{Images}

\begin{figure*}
\centerline{\psfig{figure=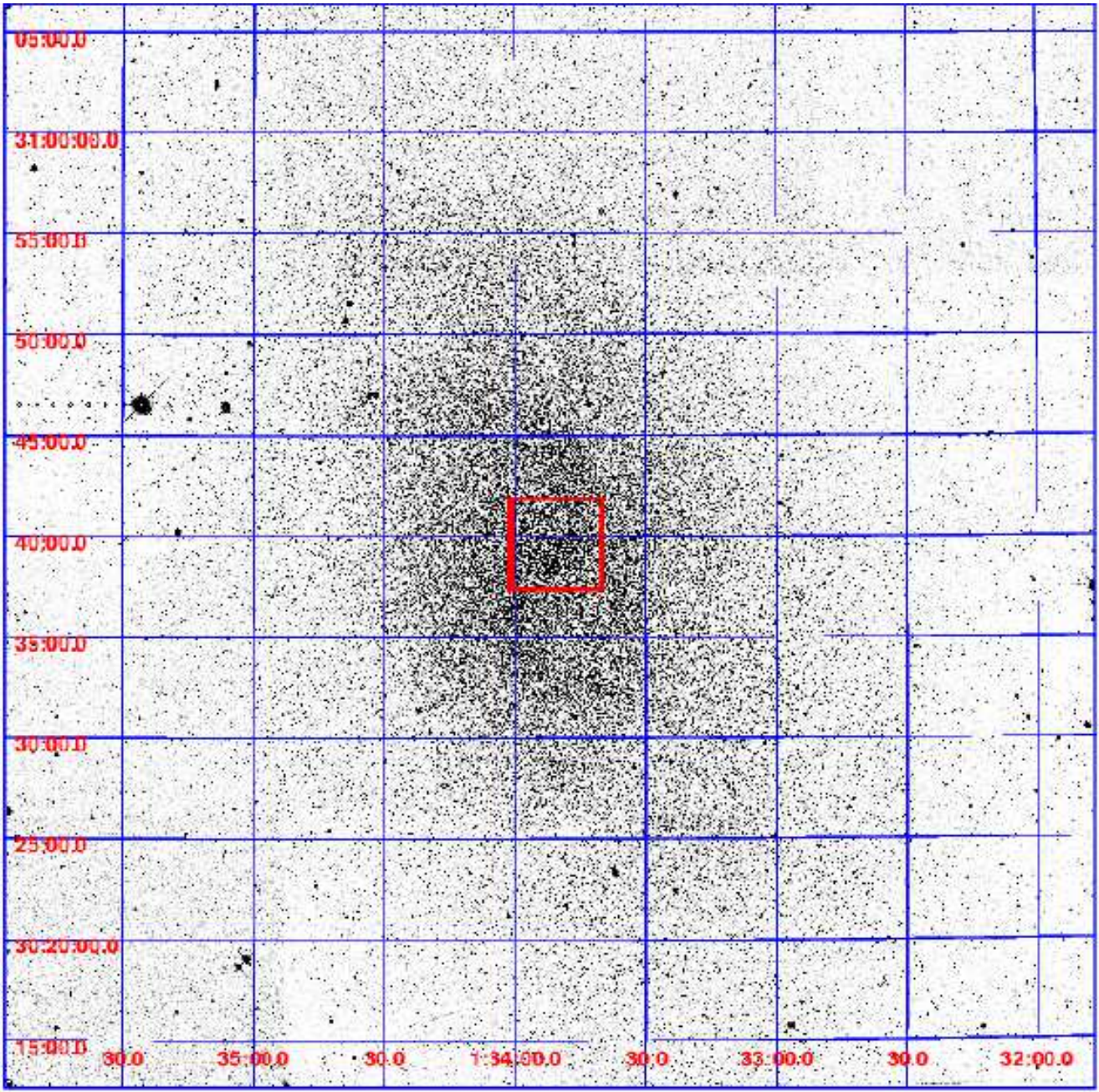,width=175mm}}
\caption[]{Combined WFCAM K-band mosaic of M\,33. The previously investigated,
square-kpc area is delineated with a box.}
\end{figure*}

We present the combined, square-degree mosaic of M\,33 in the K band in figure
1. The previously investigated, square-kpc area is delineated with a box.
While the spiral structure is evident the images are much less affected by
extinction by interstellar dust than images at optical wavelengths. The very
bright Galactic foreground red giant HD\,9687 ($1^{\rm h}35^{\rm m}26^{\rm s}$,
$+30^\circ46\rlap{.}^\prime5$) has left a trail of ghost imprints to its East,
but saturation is not normally a problem across the mosaic. Most stars above
the tip of the red giant branch (RGB) are resolved also with WFCAM, except
within the central few arcsec dominated by the nuclear star cluster.

\subsection{Cross correlation of catalogues}

The photometric catalogues of M\,33 were retrieved from the public WFCAM
Science Archive (WSA). Only the overlapping area of the other catalogues (and
images) with those from our own programme will be considered here; they are
meant to provide further epochs that increase the sensitivity and reliability
of the variability detection.

The FITS catalogues contain only un-calibrated fluxes ($f$, in counts)
obtained in a series of apertures. By knowing the photometric zeropoint at
unity airmass, $m_0$, the extinction coefficient, $a$ (Krisciunas et al.\
1987), the airmass at the start and end of observation, $z_{\rm start}$ and
$z_{\rm end}$, and exposure time, $t$, the (telluric) extinction-corrected and
flux-calibrated magnitudes, $m$, are determined by:
\begin{equation}
m=m_0-2.5\log\left(\frac{f}{t}\right)
  -a\times\left(\frac{z_{\rm start}+z_{\rm end}}{2}-1\right).
\end{equation}

In order to determine the mean magnitude and light curve of all stars over the
epochs in which they were detected, unique IDs need to be assigned to all
individual stars. To this aim, we cross-identified each star within every
catalogue. The matches were obtained by performing search iterations using
growing search radii, in steps of $0\rlap{.}^{\prime\prime}1$ out to
$1^{\prime\prime}$, on a first-encountered first-associated basis but after
ordering the principal photometry in order of diminishing brightness (to avoid
rare bright stars being erroneously associated with any of the much larger
number of faint stars).

\subsection{Photometric calibration}

Given the high level of crowding especially towards the central parts of
M\,33, and the importance of accurate relative photometry between epochs in
order to correctly separate variable from non-variable sources, it is
essential to check the accuracy of the WSA catalogues. To this aim, we
performed PSF photometry using the {\sc DAOPhot} package within IRAF (Stetson
1987) on one of the frames (from camera 1) in the central part field M\,33-3.
An empirical constant PSF model with a 2D elliptical Gaussian function was
used to construct the PSF from seven isolated stars. Then, the {\sc allstar}
task was used to estimate the instrumental magnitude for 32,627 stars
identified with the {\sc daofind} task. The transformation factor from
instrumental magnitude to standard magnitude was obtained from the standard
magnitudes of 30 stars in common from the UIST catalogue (Paper I), hence
deriving $m_0$ for the WFCAM data.

\begin{figure}
\centerline{\psfig{figure=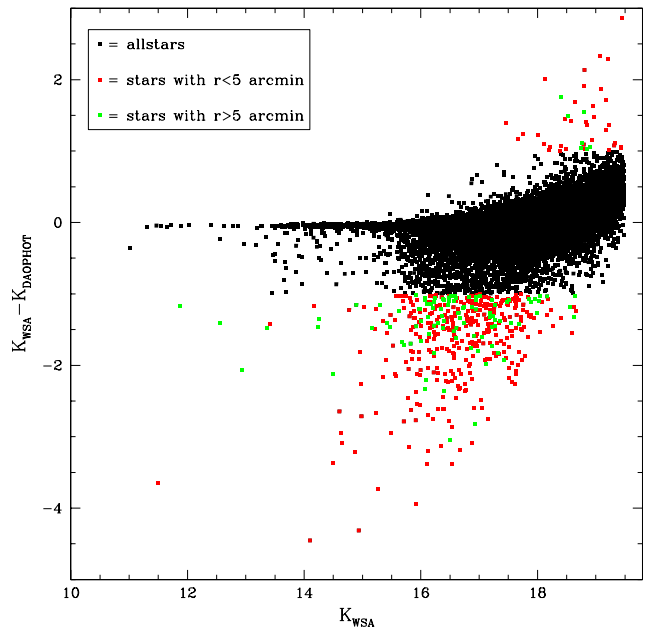,width=84mm}}
\caption[]{Magnitude differences between WSA (aperture) and {\sc DAOPhot}
(PSF) photometry, plotted against WSA magnitude. Stars for which $\Delta m>1$
mag have been identified based on whether they are located within the central
$5^\prime$ or further away from the unresolved centre of M\,33.}
\end{figure}

\begin{figure}
\centerline{\psfig{figure=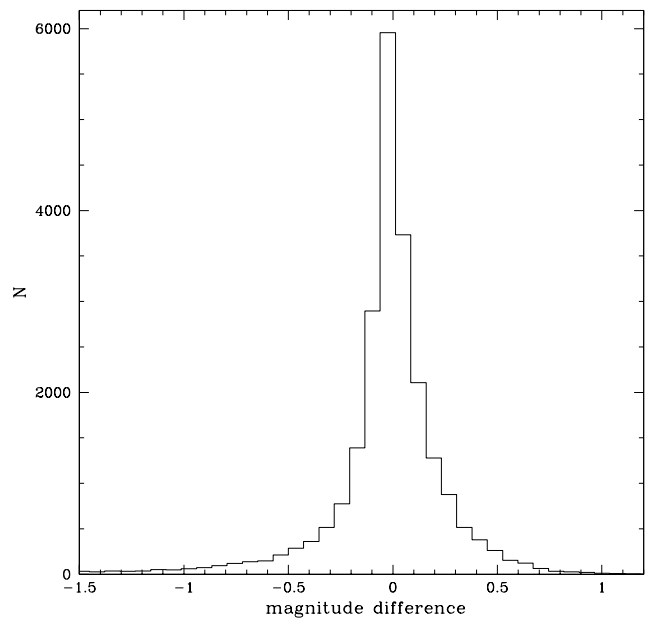,width=84mm}}
\caption[]{Histogram of the magnitude differences between WSA (aperture) and
{\sc DAOPhot} (PSF) photometry.}
\end{figure}

The celestial coordinates of the stars were calculated using the IRAF {\sc
ccmap--cctran} tasks. In this frame, the WSA catalogue lists 36\,301 stars. We
cross-identified these with the {\sc DAOPhot} catalogue within a search radius
of $1^{\prime\prime}$; hence, 24\,140 stars were identified in common. Figure 2
shows the difference in magnitude between the WSA and {\sc DAOPhot}
photometry, against the WSA magnitude; the histogram of magnitude differences
is shown in figure 3. The vast majority of stars have magnitudes that are
consistent between the two methods of measurement within a few tenths of a
magnitude; among the 3.5 per cent of stars with $\Delta m>1$ mag, 78 per cent
are located near (within $5^\prime$ of) the centre of M\,33 where crowding is
more severe than elsewhere in the mosaic. This renders just 0.9 per cent of
stars with suspect photometry. Upon visual inspection of the image, it was
found that in most cases there is a faint star near a bright star, and the
photometric difference is simply the result of erroneous cross-identification.
Consequently, the accuracy of the WSA magnitudes is acceptable.

\subsubsection{Relative calibration}
 
\begin{figure}
\centerline{\psfig{figure=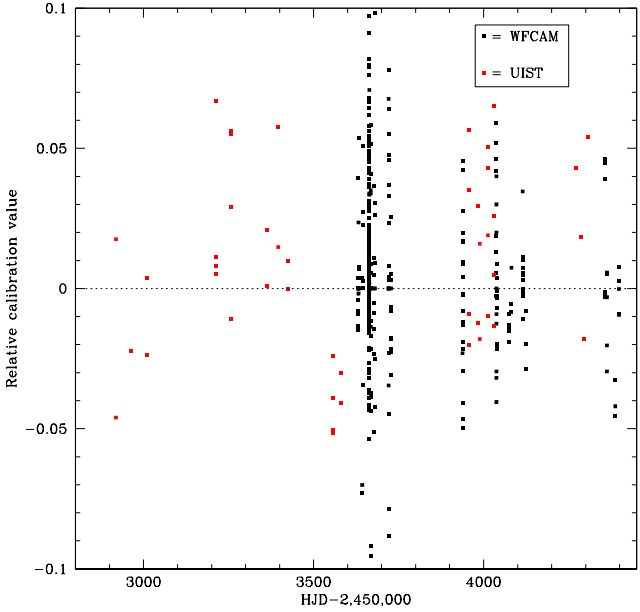,width=84mm}}
\caption[]{Relative calibration values added to the magnitudes in each of the
individual frames, as a function of time, over the duration of the UIST and
WFCAM monitoring campaigns.}
\end{figure}

The relative calibration between frames was obtained from the mean magnitudes
of $\approx1000$ stars in common within the magnitude interval $K\in
[16...18]$. While these will include some variable stars the vast majority
will not vary by more than 10 per cent rendering their mean magnitude accurate
to well within a per cent. Then the photometry from the different frames was
brought in line with each other by applying corrections that equalised these
mean magnitudes. The corrections are shown in figure 4, also for the UIST
survey (they were not shown in Paper I). Note that the corrections are small
-- generally a few per cent and always less than 10 per cent.

\section{Variability analysis}

The search for variable stars was done by calculating the variability index
for each star. This index was introduced by Welsh \& Stetson (1993) and
developed further by Stetson (1996). In this method, first the observations
are paired on the basis of timespan between observations such that the
observations of each pair have a timespan less than the shortest period
expected for the kind of variable star of interest. In case more than two
observations were performed within the timespan of shortest periodicity,
those sets of observations would be paired in more than one pair. Hence, the
$J$ index is calculated:
\begin{equation}
J=\frac{\Sigma_{k=1}^N w_k\ {\rm sign}(P_k)\sqrt{|P_k|}}{\Sigma_{k=1}^N w_k}.
\end{equation}
Here, observations $i$ and $j$ have been paired and for each pair a weight
$w_k$ is assigned; the product of normalized residuals,
$P_k=(\delta_i\delta_j)_k$\footnote{Following Stetson (1996), $P_k=\delta^2-1$
if $i=j$.}, where $\delta_i=(m_i-\langle m\rangle)/\epsilon_i)$, is the
residual of measurement $i$ from the mean magnitude, normalized by the error
of the measurement, $\epsilon_i$; and $N$ is the total number of observations.
Note that $\delta_i$ and $\delta_j$ may refer to observations taken in
different filters. The $J$ index for non-variable stars is approximately zero
as the residuals arising from random noise are uncorrelated and their product
will therefore tend to zero for large sets of measurements.

The effect of a small number of observations or corrupt data can be limited by
means of a backup index, viz.\ the Kurtosis index:
\begin{equation}
K=\frac{\frac{1}{N} \Sigma_{i=1}^n |\delta_i|}{\sqrt{\frac{1}{N}\Sigma_{i=1}^N
\delta_i^2}}.
\end{equation}
The shape of the light variations define the value of the Kurtosis index; for
example, $K=0.798$ for a Gaussian distribution which is concentrated towards
the average brightness level (as would be random noise), and $K\rightarrow0$
for data affected by a single outlier (when $N\rightarrow\infty$).

Here, we use the variability index $L$ (Stetson 1996):
\begin{equation}
L=\frac{J\times K}{0.798} \frac {\Sigma w}{w_{\rm all}},
\end{equation}
where $\Sigma w$ is the total weight assigned to a given star and $w_{\rm all}$
is the total weight a star would have if observed in every single observation.

\begin{figure}
\centerline{\psfig{figure=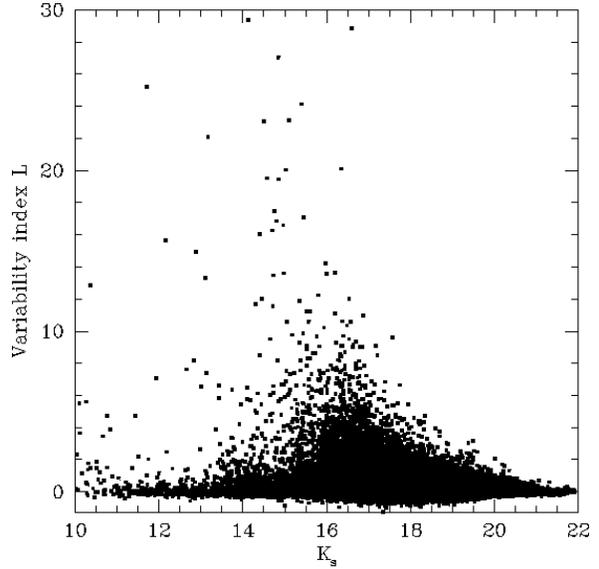,width=84mm}}
\caption[]{Variability index $L$ {\it vs.}\ K-band magnitude.}
\end{figure}

\begin{figure}
\centerline{\psfig{figure=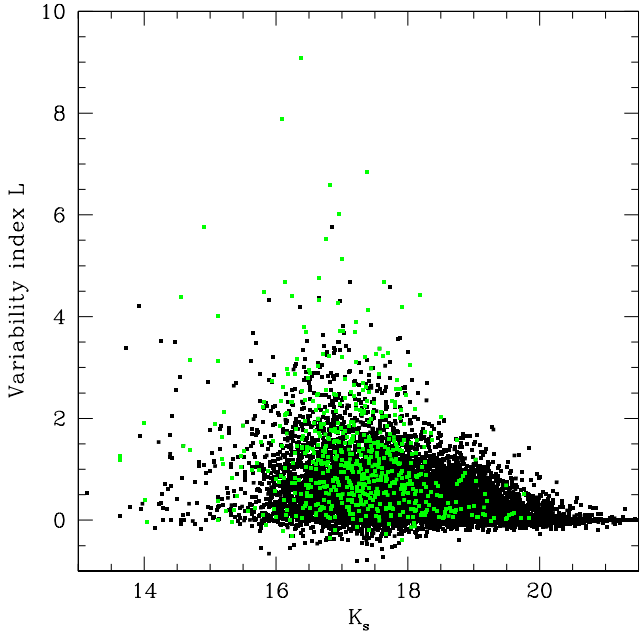,width=84mm}}
\caption[]{Variability index $L$ {\it vs.}\ K-band magnitude (from the WFCAM
data) for the central square kpc field that was monitored with UIST (Paper I).
The green points are the UIST variable stars that were detected with WFCAM.} 
\end{figure}

Figure 5 shows how the variability index $L$ varies with K-band magnitude. For
comparison with our previous UIST survey (Paper I), in figure 6 we show the
distribution for the central square kpc of M\,33, where we indicate the WFCAM
detections of stars that had been identified as variable in the UIST survey.
Both graphs reveal a noticeable ``branch'' of stars with larger than usual $L$
between $K\sim16$--18 mag; these are likely AGB stars with Mira-type
variability.

\begin{figure}
\centerline{\psfig{figure=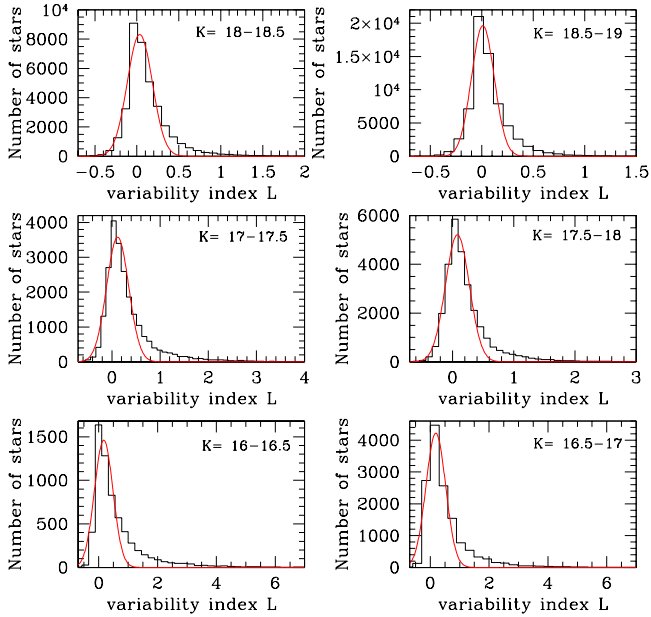,width=84mm}}
\caption[]{Variability index $L$ histograms for several magnitudes bins in the
range $K_{\rm s}=16$--19 mag. Red lines trace the Gaussian function fitted to
each histogram.}
\end{figure}

Several tests were performed to select the optimal variability index
threshold. First we inspected the histograms of variability index within
several magnitude bins in the range 16--19 mag for all detected stars
including the M\,33 disc and central regions (Fig.\ 7). To determine the
variability index threshold, a Gaussian function was fitted to each of these
histograms. The Gaussian function should be near-perfectly fitted to those
data for low values of $L$, while it departs from the histograms for larger
values of $L$. This appears to happen around $L\sim0.8$ but at smaller values
for faint stars; we thus decided, in first instance, to set the threshold at
$L>0.7$ for the detection of variability.

\begin{figure}
\centerline{\psfig{figure=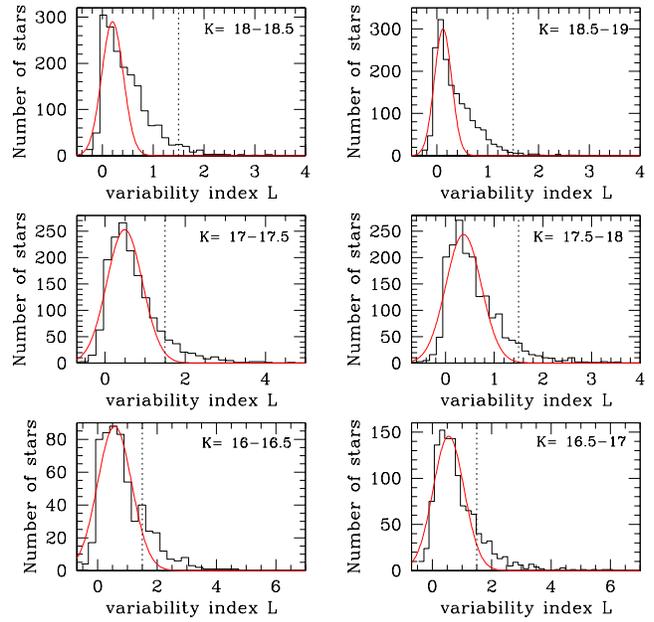,width=84mm}}
\caption[]{As Fig.\ 7, but limited to the central square kpc of M\,33. The
vertical dotted line marks the limit which is used for selection of variable
stars in this part of M\,33.}
\end{figure}

\begin{figure}
\centerline{\psfig{figure=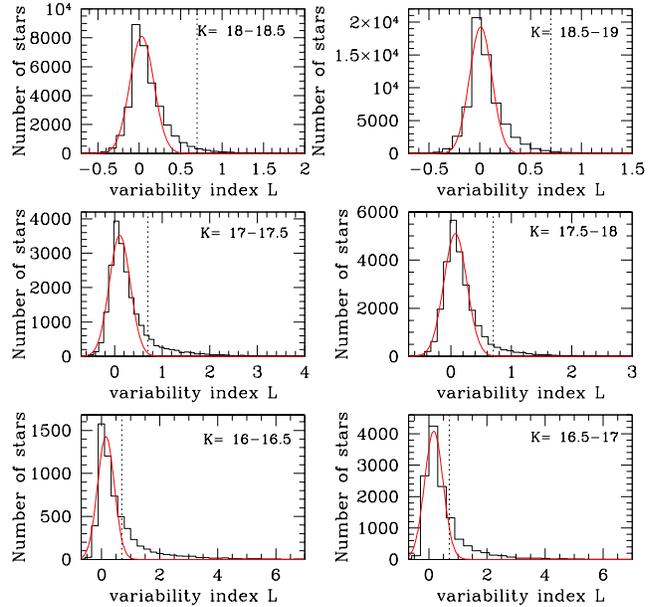,width=84mm}}
\caption[]{As Fig.\ 8, but now excluding the central square kpc.}
\end{figure}

Next, we test our procedure by comparing the WFCAM selected variables, with
$L>0.7$, with the UIST catalogue of variables in the central square kpc (from
Paper I). The percentage of WFCAM variables within the magnitude ranges shown
in figure 7 is about twice that derived from the UIST survey. If anything,
the opposite would be expected as the UIST survey had a superior cadence of
observations. Indeed, when selecting only stars from this central region of
M\,33, the histograms of variability index broaden, and a more appropriate
threshold would be $L>1.5$ (Fig.\ 8). On the other hand, a threshold of
$L>0.7$ seems appropriate for the disc of M\,33, i.e.\ excluding the central
square kpc (Fig.\ 9).

\begin{figure}
\centerline{\psfig{figure=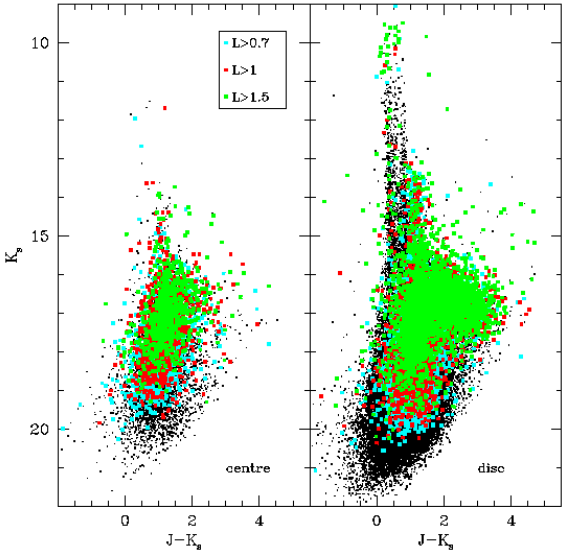,width=84mm}}
\caption[]{Near-IR colour--magnitude diagram of M\,33 with indicated the
putative variable stars according to three choices of variability index $L$
threshold, for ({\it Left:}) the central square kpc and ({\it Right:}) the
disc.}
\end{figure}

To further assess the validity of the choice of variability index threshold,
we examined the location of the putative variable stars in a colour--magnitude
diagram (CMD), for different choices of $L$ threshold: $L>0.7$, $L>1$ and
$L>1.5$ (Fig.\ 10). While there is no clear difference in the selection of
variable stars on the AGB and RSG portions of the CMD (roughly at $K_{\rm
s}=14$--18.5 mag and $(J-L_{\rm s})>0.8$ mag), all but the $L>1.5$ thresholds
result in many putative variable stars at fainter magnitudes and along the
blue and bright vertical sequence (around $(J-K_{\rm s})\sim0.5$ mag) where no
large-amplitude red giant variables are expected.

\begin{figure}
\centerline{\psfig{figure=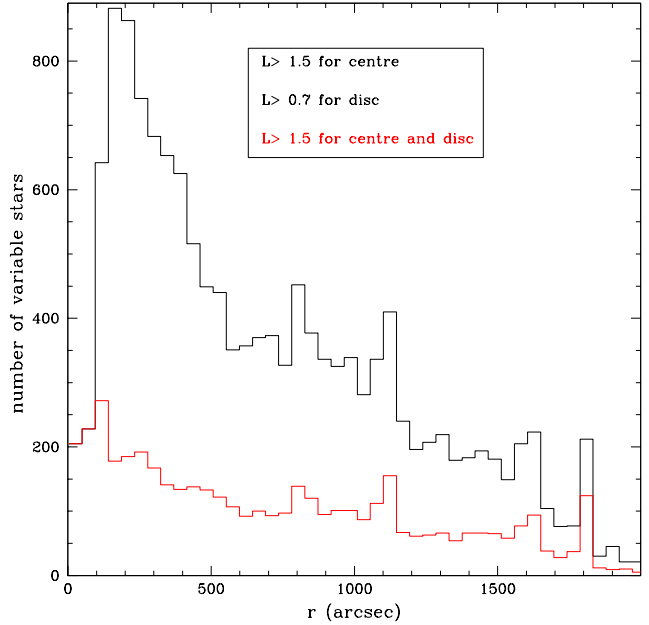,width=84mm}}
\caption[]{Radial distribution of putative variable stars where ({\it Left:})
different variability index $L$ thresholds were applied for the central square
kpc and the disc and ({\it Right:}) the same thresholds were applied.}
\end{figure}

Finally, we inspected the radial distribution of putative variable stars for
the case where we apply different choices for the $L$ threshold for the
central square kpc and the (remainder of the) disc, as compared to applying a
single threshold for all (Fig.\ 11). The former case results in a break in
gradient around $r\sim500^{\prime\prime}$ -- i.e.\ well outside the central
square kpc -- where $L>0.7$ results in a seemingly disproportionally increased
number of variable stars closer to the centre. This gradient is not sustained
within the central $r\sim150^{\prime\prime}$ -- i.e.\ corresponding roughly to
the central square kpc. Indeed, the distribution turns over and much fewer
variable stars are identified once $L>1.5$ was applied. The same artificial
behaviour is not seen when the same $L$ threshold is applied across the entire
M\,33 galaxy.

Based on the histograms, CMD and radial distributions we decided to apply a
single variability threshold of $L>1.5$.

\subsection{Comparison between the WFCAM and UIST catalogues of variable stars
within the central square kpc}

\begin{figure}
\centerline{\psfig{figure=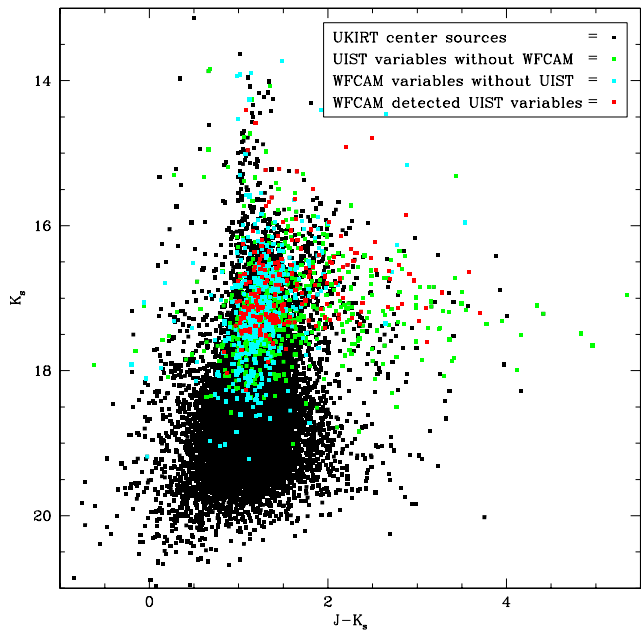,width=84mm}}
\caption[]{Near-IR CMD of the central square kpc of M\,33, showing those stars
from the UKIRT/WFCAM survey that were and were not detected in the UKIRT/UIST
survey (Paper I). A variability threshold of $L>1.5$ was applied to select
variable stars from the WFCAM survey.}
\end{figure}

\begin{figure}
\centerline{\psfig{figure=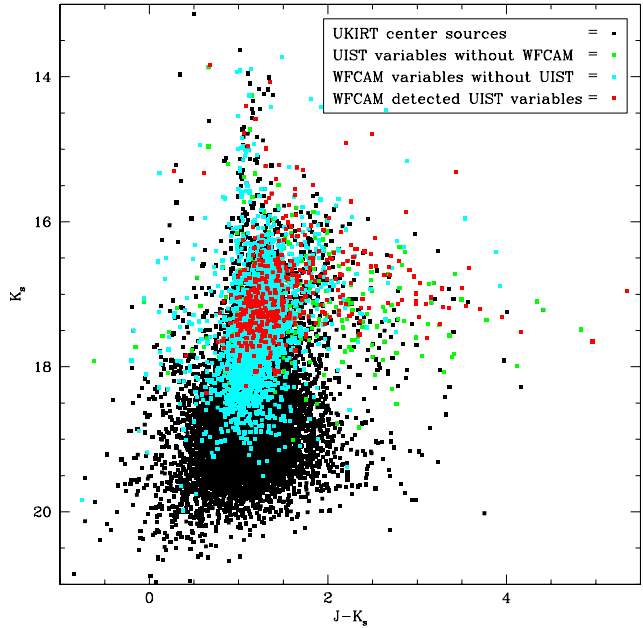,width=84mm}}
\caption[]{Same as Fig.\ 12, but for a threshold of $L>0.7$.}
\end{figure}

One final assessment of the variability detection success is made by a more
careful comparison of the common area of the WFCAM and UIST surveys. After
sorting both catalogues in order of diminishing brightness in K band, matches
were obtained through successive search iterations using increasing search
radii, in steps of $0\rlap{.}^{\prime\prime}1$ out to $1^{\prime\prime}$. Within
the coverage of UIST (Paper I), using WFCAM we detected 11\,114 stars; from
18\,398 stars detected with UIST, using WFCAM we detected 10\,095 stars.
Applying $L>1.5$, we find 667 variable stars located within the central square
kpc covered by UIST (out of 4643 in total across the WFCAM coverage); applying
$L>0.7$, this becomes 2696 -- i.e.\ more than three times as many as were
found in the more capable UIST survey. Again applying $L>1.5$, only 192 out of
the 812 UIST LPVs were identified with WFCAM (a success rate of 24\%). This
also suggests that the UIST survey, too, is incomplete.

Figures 12 and 13 show CMDs in which are highlighted those variable stars
which have and those which have not been detected with either WFCAM of UIST,
when applying a WFCAM variability index threshold of $L>1.5$ and $L>0.7$,
respectively. These CMDs suggest that $L>1.5$ is a more sensible threshold
than $L>0.7$, for two reasons; firstly, it yields a somewhat smaller number of
variables with WFCAM than with UIST, which is expected as WFCAM has more
difficulty in isolating blended stars and also the WFCAM survey is based on
fewer epochs. Secondly, the fraction of UIST variables that were found with
WFCAM is smaller but the fraction of WFCAM variables that were found with UIST
is much larger (31\%) than when applying $L>0.7$ (14\%).

\begin{figure}
\centerline{\psfig{figure=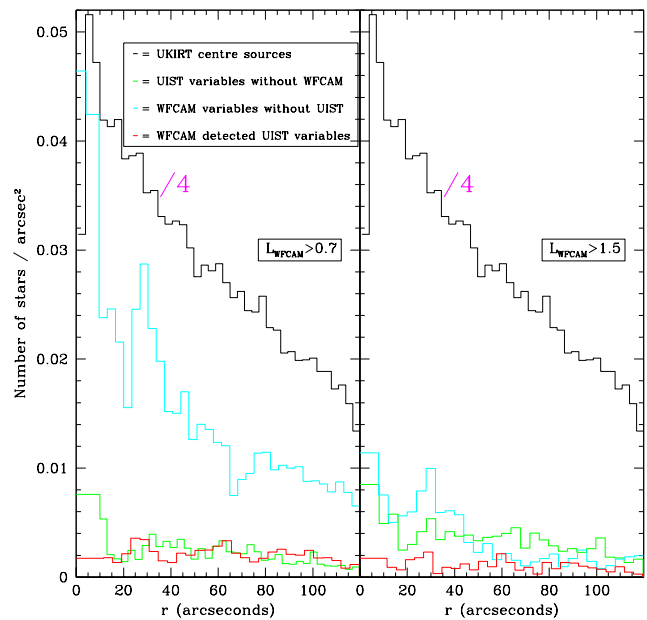,width=84mm}}
\caption[]{Radial profiles of total stellar density (divided by four for ease
of comparison) and density of variable stars found in the WFCAM and/or UIST
surveys, applying a variability index threshold of $L>0.7$ ({\it Left}) and
$L>1.5$ ({\it Right}), respectively.}
\end{figure}

\begin{figure}
\centerline{\psfig{figure=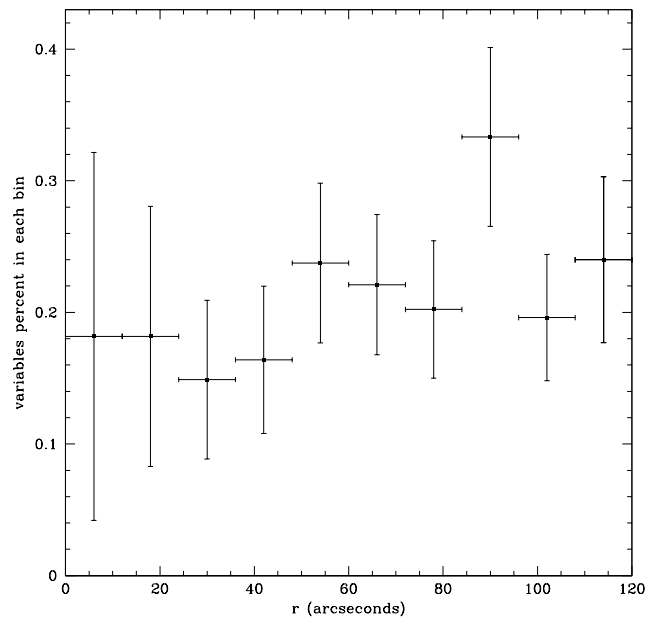,width=84mm}}
\caption[]{Radial profile of the fraction of UIST variables that is recovered
in the WFCAM variability survey.}
\end{figure}

Since the central part of M\,33 comprises a considerable variation in stellar
density, the same data displayed in figures 12 and 13 are shown in figure 14
as a function of radial distance from the centre of M\,33. The rate at which
variable stars are detected increases towards the centre, and as a consequence
the number of variable stars found in one of the surveys that is not recovered
within the other survey also increases towards the centre. However, when
applying a variability index threshold of $L_{\rm WFCAM}>0.7$, the WFCAM survey
yields many more variable stars than the UIST survey, whilst for
$L_{\rm WFCAM}>1.5$ this is about equal. This again favours the choice of the
latter threshold. The success rate of the WFCAM survey to recover UIST
variable stars displays only a shallow gradient with radial distance from the
centre of M\,33 (Fig.\ 15); it averages $\sim22$\% for this central region
(but is $\sim24$\% for the square area encompassing this circular area).

\begin{figure}
\centerline{\psfig{figure=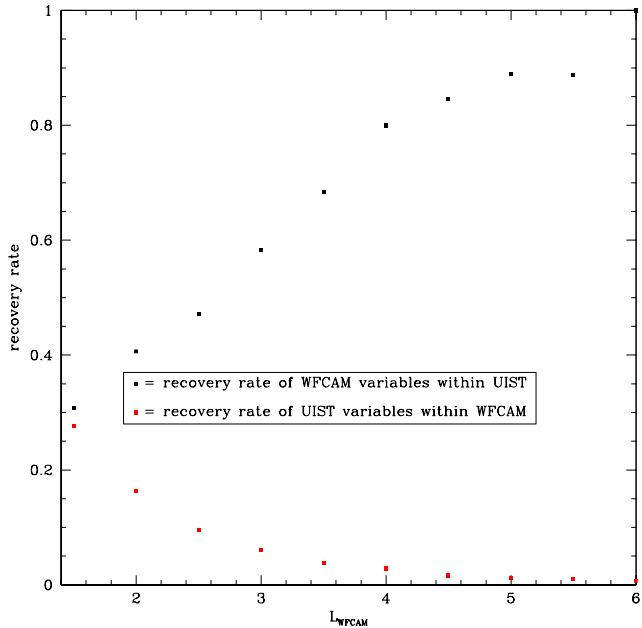,width=84mm}}
\caption[]{Recovery rate of UIST variable stars within the WFCAM survey, and
vice versa, as a function of WFCAM variability index threshold ($L$).}
\end{figure}

The recovery rate of UIST variables within the WFCAM catalogue of variable
stars, and vice versa, is plotted as a function of WFCAM variability index
threshold in figure 16. At $L_{\rm WFCAM}>1.5$, a near-equal fraction of variable
stars is recovered within each of the surveys ($\sim30$\%) -- the UIST survey
is slightly more successful than the WFCAM survey, as expected (see above). As
$L_{\rm WFCAM}$ increases, the number of variable stars identified in the WFCAM
survey decreases but these variable stars will be the ones with larger
amplitudes. Hence, the UIST survey will be more complete in including those
WFCAM variable stars, reaching $\sim90$\% success rate for $L_{\rm WFCAM}>5$. On
the other hand, the WFCAM survey will miss more of the UIST variable stars
(generally the ones with smaller amplitudes), and its success rate drops to
just one per cent for $L_{\rm WFCAM}>5$. Hence, the choice for $L_{\rm WFCAM}>1.5$
is supported once again.

\subsection{Amplitudes of variability}

We estimate the amplitude of variability by assuming a sinusoidal lightcurve
shape. The amplitude is then:
\begin{equation}
A\approx2\times\sigma/0.701
\end{equation}
where $\sigma$ is the standard deviation in our data and 0.701 is the standard 
deviation of a unit sine function. The standard deviation is uncertain when
the number of measurements ($N$) is small; we showed in Paper I that $N=6$ is
the minimum acceptable to reach $\sim10$\% fidelity.

\begin{figure}
\centerline{\psfig{figure=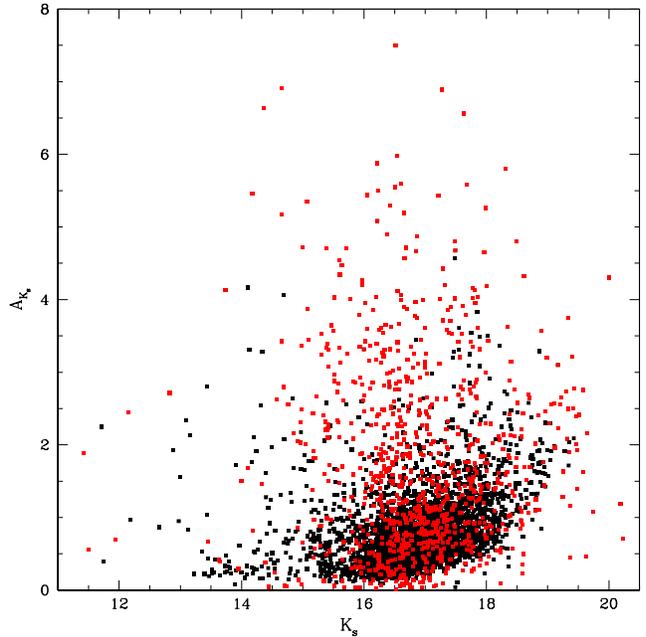,width=84mm}}
\caption[]{Estimated amplitude, $A_{\rm K}$, of variability {\it vs.}\ K-band
magnitude. Stars with $\leq6$ measurements are highlighted in red.}
\end{figure}

The estimated K-band amplitude is plotted {\it vs}.\ K-band magnitude in
figure 17. As is well known by now (Wood et al.\ 1992; Wood 1998; Whitelock et
al.\ 2003; Paper I), the amplitude of variability exhibits a clear tendency to
diminish with increasing brightness, which is partly due to the definition of
magnitude as a relative measure rather than less powerful pulsation (van Loon
et al.\ 2008). The amplitude is generally about a magnitude or less, but a
small fraction of variable stars (8\%) seem to reach $A_K>2$ mag. Very dusty
AGB stars -- which are rare -- are known to reach such large amplitudes (Wood
et al.\ 1992; Wood 1998; Whitelock et al.\ 2003). Among these extreme
variables, 311 stars have $N\leq6$; excluding these stars, only 20 stars
remain with $A_K>3$ mag. The brightest variables, with $K_{\rm s}<13$ mag, are
foreground red giants; among the faintest stars, with $K_{\rm s}>17$ mag, our
survey becomes increasingly less sensitive to small-amplitude variables.

\section{Description of the catalogue}

\begin{table}
\caption[]{Description of the photometric catalogue.}
\begin{tabular}{ll}
\hline\hline
Column No. & Descriptor                    \\
\hline
\multicolumn{2}{l}{\it Part I: stellar mean properties (403,734 lines)} \\
 1         & Star number                   \\
 2         & Right Ascension (J2000)       \\
 3         & Declination (J2000)           \\
 4         & Mean J-band magnitude         \\
 5         & Error in $\langle J\rangle$   \\
 6         & Mean H-band magnitude         \\
 7         & Error in $\langle H\rangle$   \\
 8         & Mean K-band magnitude         \\
 9         & Error in $\langle K\rangle$   \\
10         & Number of J-band measurements \\
11         & Number of H-band measurements \\
12         & Number of K-band measurements \\
13         & Variability index $J$         \\
14         & Kurtosis index $K$            \\
15         & Variability index $L$         \\
16         & Estimated K-band amplitude    \\
\multicolumn{2}{l}{\it Part II: multi-epoch data (3,623,332 lines)} \\
 1         & Star number                   \\
 2         & Epoch (HJD--2,450,000)        \\
 3         & Filter (J, H or K)            \\
 4         & Magnitude                     \\
 5         & Error in magnitude            \\
\hline
\end{tabular}
\end{table}

The photometric catalogue including all variable and non-variable stars is
made publicly available at the Centre de Donn\'ees astronomiques de Strasbourg
(CDS). The content is described in Table 4. It is composed of two parts, part
I comprising the mean properties of the stars and part II tabulating all the
photometry (for the benefit of generating lightcurves, for instance).

The astrometric accuracy of the catalogue is $\approx0.2^{\prime\prime}$ r.m.s.,
tied to the 2MASS system. This accuracy was found to be consistent with the
results from our cross-correlations with three other optical and IR catalogues
(cf.\ Section 5.3).

\section{Discussion}

\subsection{The near-IR variable star population}

\begin{figure}
\centerline{\psfig{figure=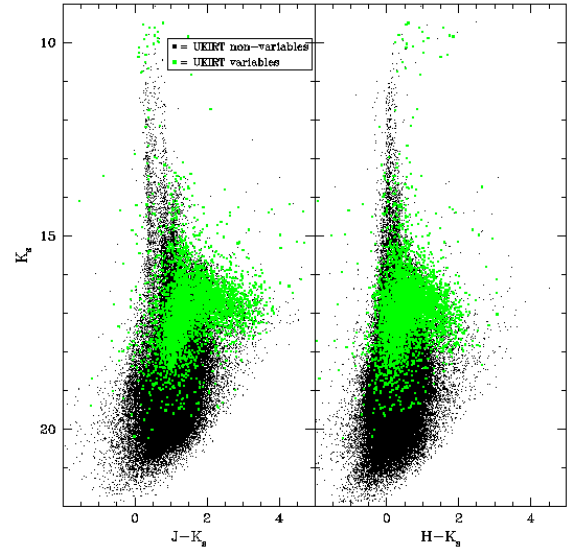,width=84mm}}
\caption[]{Near-IR CMDs (WFCAM variable stars in green).}
\end{figure}

Figure 18 presents near-IR CMDs for the whole region of M\,33 monitored with
WFCAM. The Large-amplitude variable stars we identified are highlighted in
green. These are mainly found between $K_{\rm s}\sim16$--18 mag, and are
largely absent among fainter stars (below the tip of the RGB  at
$K_{\rm s}\sim18$ mag). Some brighter variable RSGs are found (around
$K_{\rm s}\sim14$ mag), but the clump of variables stars with $K_{\rm s}<11$
mag and redder colour in $H-K_{\rm s}$ than $J-K_{\rm s}$ are foreground stars
and perhaps saturated. Variable stars dominate the redder stars to the right
of the vertical sequence comprising the bulk of stars.

\begin{figure}
\centerline{\psfig{figure=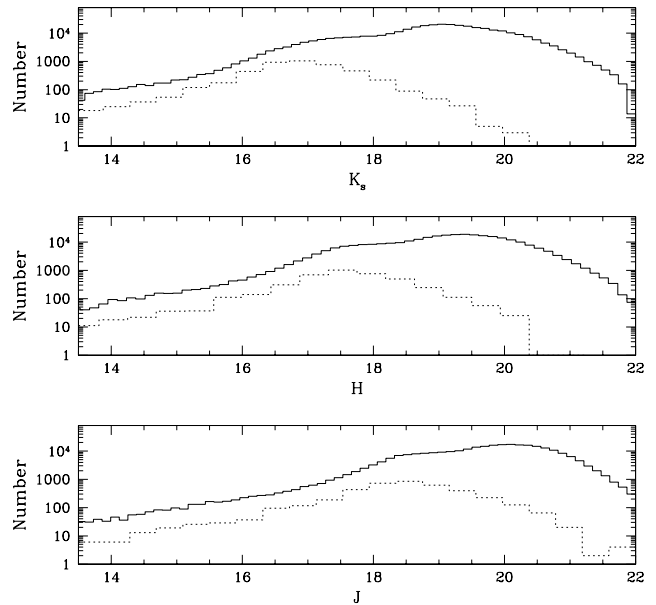,width=84mm}}
\caption[]{Distribution of all WFCAM sources (solid) and the variable stars
(dotted), as a function of near-IR brightness.}
\end{figure}

\begin{figure}
\centerline{\psfig{figure=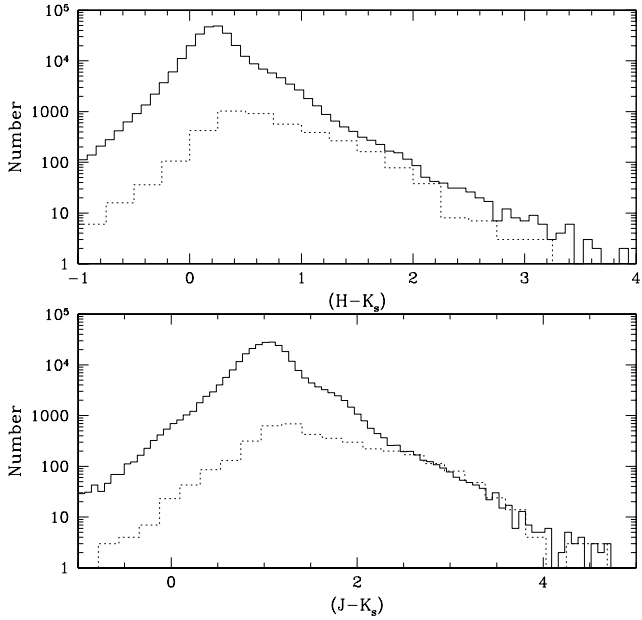,width=84mm}}
\caption[]{Distribution of all WFCAM sources with $K_{\rm s}<19$ mag (solid)
and the variable stars (dotted), as a function of near-IR colour.}
\end{figure}

The distributions over brightness (Fig.\ 19) and colour (Fig.\ 20) provide
another means of assessing the properties of the variable stars. The largest
fraction of stars that are found to be variable occurs between
$K_{\rm s}\sim16$--17 mag, albeit less than that of the UIST survey in the
central region in the same magnitude interval. The variable star population
reaches a peak around $K_{\rm s}\sim17$ mag; it then drops at fainter
magnitudes even though the total stellar population keeps increasing. This
mainly arises from two factors: firstly, many stars in this magnitude interval
have not yet reached the final phase of their evolution and they will still
evolve to higher luminosities and lower temperatures before they develop
large-amplitude variability; secondly, the birth-mass and K-band brightness
relation flattens considerably for low-mas AGB stars (see Paper II). At
$(J-K_{\rm s})>2$ mag or $(H-K_{\rm s})>1$ mag almost all stars ($K_{\rm s}<19$
mag) are variable; these are dusty, strongly pulsating and heavily mass-losing
AGB stars.

\begin{figure}
\centerline{\psfig{figure=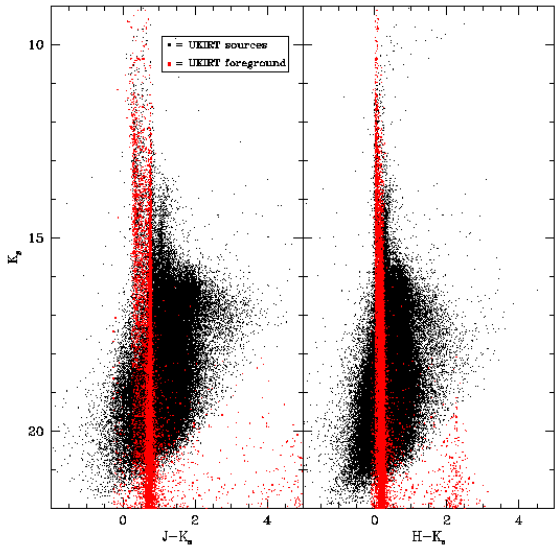,width=84mm}}
\caption[]{Estimated contamination by foreground stars (in red), from a
simulation with {\sc trilegal} (Girardi et al.\ 2005).}
\end{figure}

The level of contamination by foreground stars can be assessed with the {\sc
trilegal} simulation tool (Girardi et al.\ 2005). We used default parameters
for the structure of the Galaxy, simulating a 0.9 square degree field in the
direction ($l=133.61^\circ$, $b=-31.33^\circ$). Only a relatively small number
of foreground stars are expected, with fairly neutral colours or below our
completeness limit (Fig.\ 21). The part of the CMD occupied by large-amplitude
variable stars is relatively uncontaminated by foreground stars.

\begin{figure}
\centerline{\psfig{figure=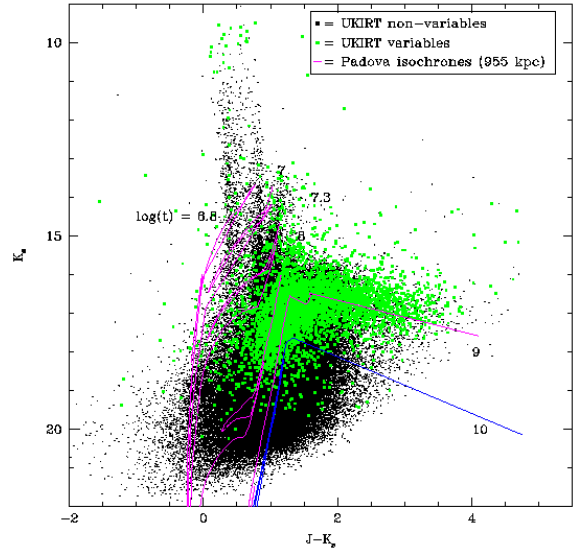,width=84mm}}
\caption[]{CMD of $(J-K_{\rm s})$, with WFCAM variable stars in green.
Overplotted are isochrones from Marigo et al.\ (2008) for solar metallicity
and a distance modulus of $\mu=24.9$ mag.}
\end{figure}

The stellar populations can be described using isochrones calculated by Marigo
et al.\ (2008) (Fig.\ 22). The isochrones were calculated for solar
metallicity ($Z_\odot=0.015$) for all stellar populations except the oldest,
least chemically evolved one with $\log t[{\rm yr}]=10$ for which we adopted
$Z=0.008$. To account for a (shallow) metallicity gradient across the disc of
M\,33, we show CMDs of each of the 16 tiles of the M\,33 mosaic, overlain with
isochrones for $Z=0.015$ in the central region but $Z=0.008$ further out in
the disc (Fig.\ 23).

\begin{figure*}
\centerline{\psfig{figure=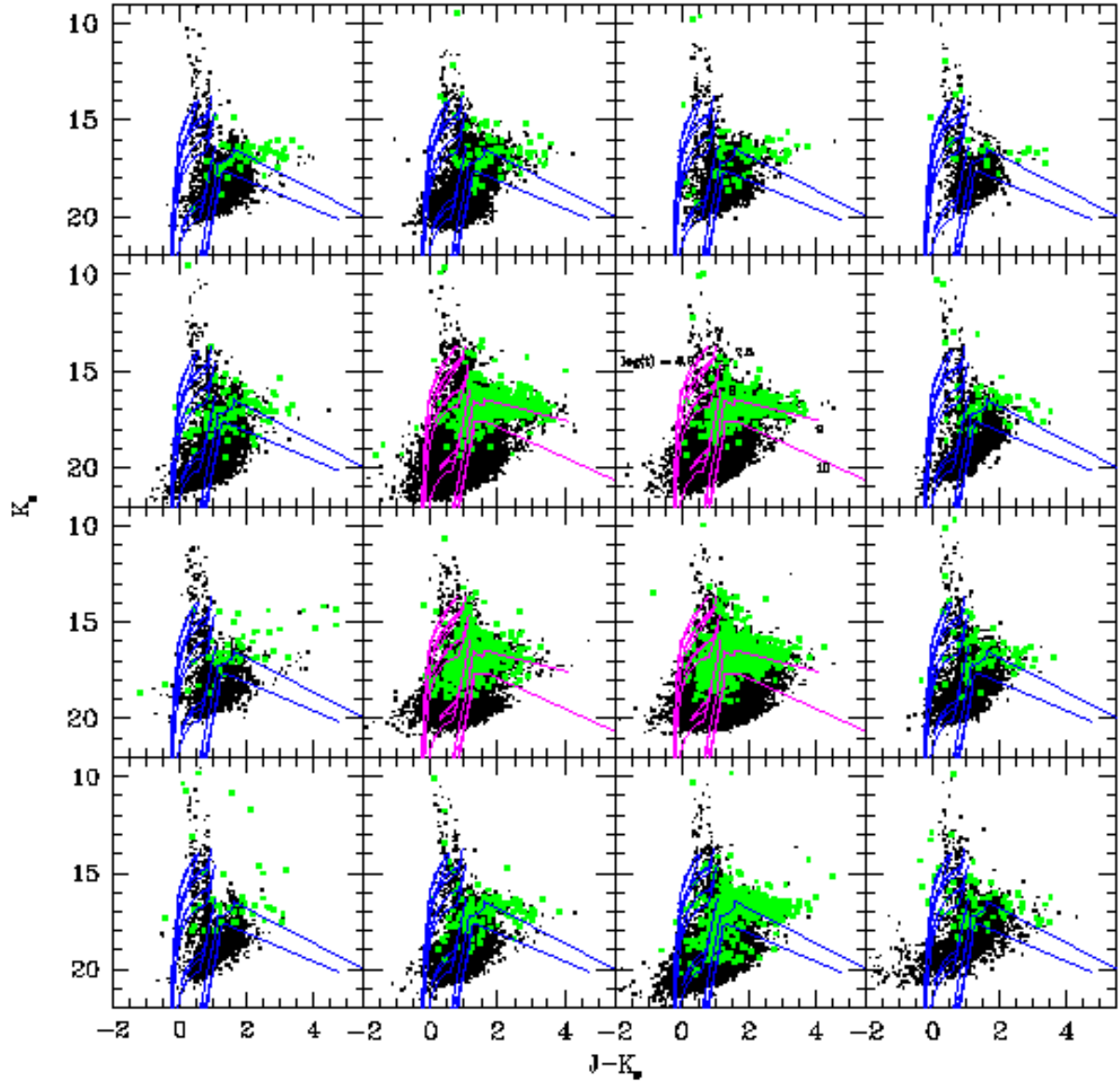,width=180mm}}
\caption[]{Near-IR CMDs of each of the individual tiles in our mosaic. The
WFCAM variable stars are highlighted in green. Isochrones from Marigo et al.\
(2008) are overlain for $Z=0.015$ (pink) and $Z=0.008$ (blue).}
\end{figure*}

The isochrones from Marigo et al. (2008) are the most realistic models to be
used for the purpose of this study, for the folowing reasons:
\begin{itemize}
\item[$\bullet$]{The star's evolution is followed all the way through the
thermal pulsing AGB until the post-AGB phase. Crucially, two important phases
of stellar evolution are included, viz.\ the third dredge-up mixing of the
stellar mantle as a result of the helium-burning phase, and the enhanced
luminosity of massive AGB stars undergoing hot bottom burning (HBB; Iben \&
Renzini 1983);}
\item[$\bullet$]{The molecular opacities which are important for the cool
atmospheres of evolved stars have been considered in the models of stellar
structure. The transformation from oxygen-dominated (M-type) AGB stars to
carbon stars in the birth mass range $M\sim1.5$--4 M$_\odot$ is accounted for
(cf.\ Girardi \& Marigo 2007);}
\item[$\bullet$]{The dust production in the winds of LPVs, and the associated
reddening is included;}
\item[$\bullet$]{The radial pulsation mode is predicted;}
\item[$\bullet$]{Combination of their own models for intermediate-mass stars
($M<7$ M$_\odot$), with Padova models for more massive stars  ($M>7$ M$_\odot$;
Bertelli et al.\ 1994), gives a complete coverage in birth mass ($0.8<M<30$
M$_\odot$);}
\item[$\bullet$]{Magnitudes are calculated on a wide range of common optical
and IR photometric systems;}
\item[$\bullet$]{The isochrones are available via an internet-based form, in a
user-friendly format.}
\end{itemize}

\subsection{Spatial distribution of LPVs}

\begin{figure*}
\hbox{
\psfig{figure=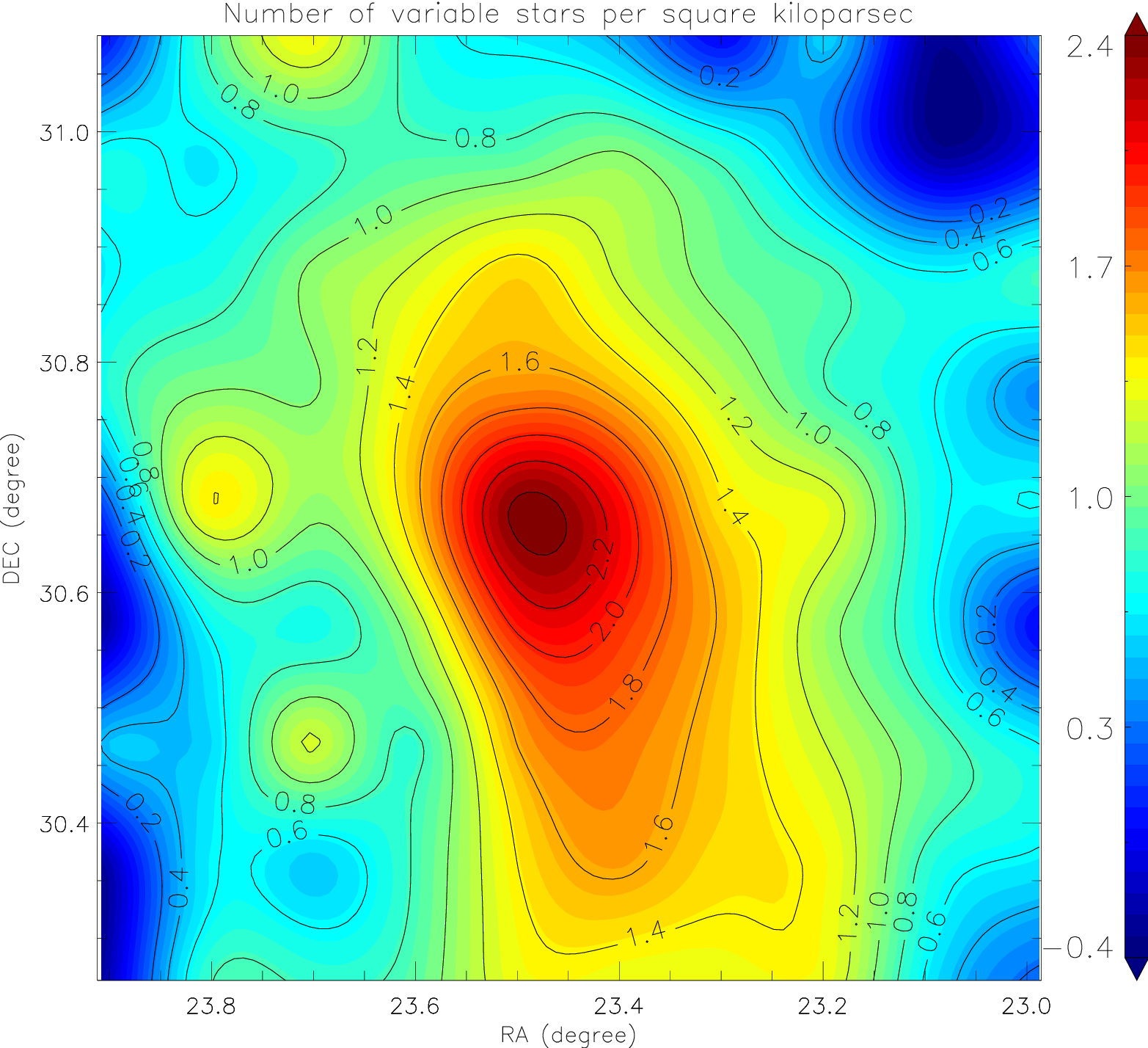,width=85mm}
\hspace*{3mm}
\psfig{figure=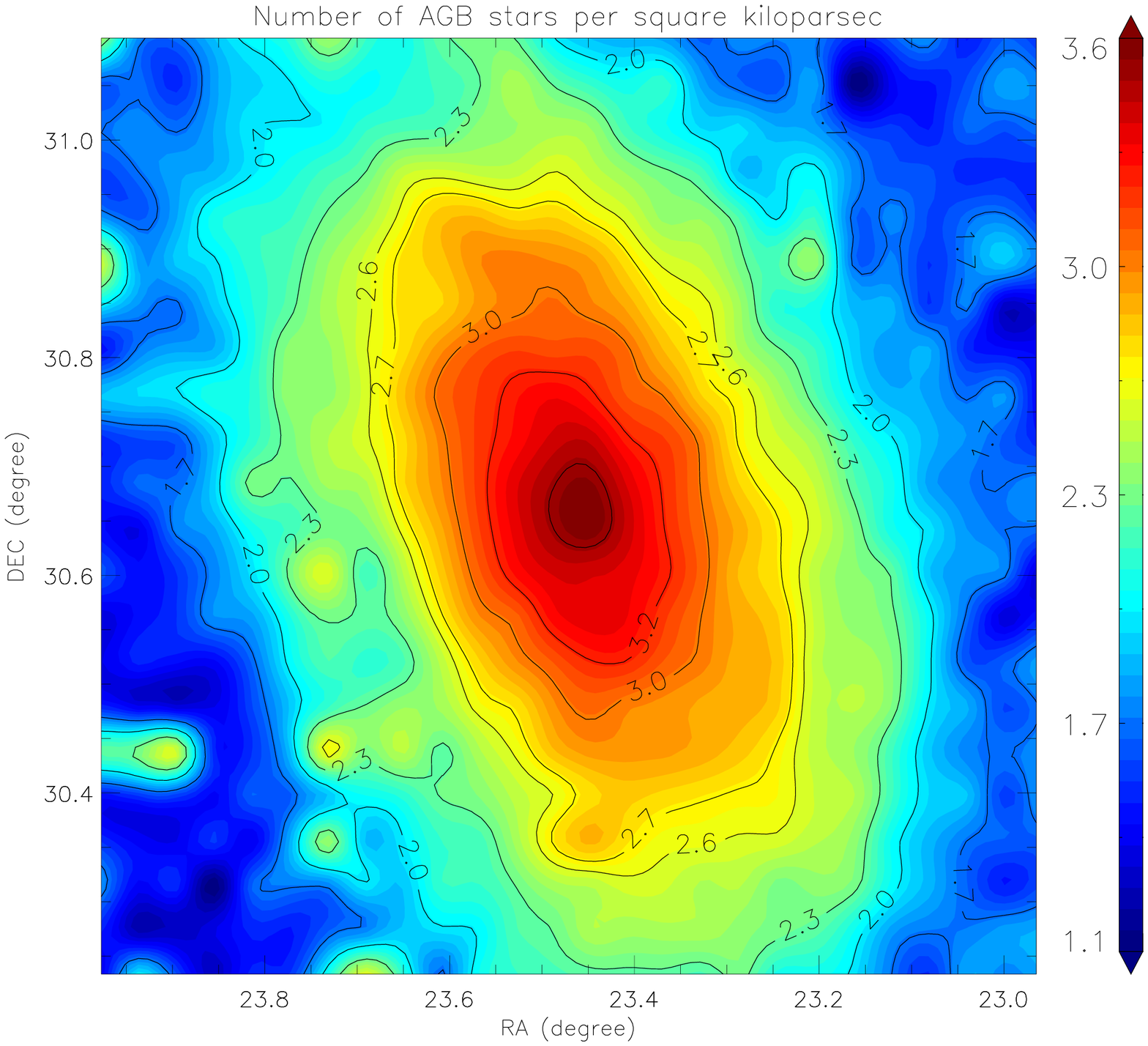,width=85mm}
}
\hbox{
\psfig{figure=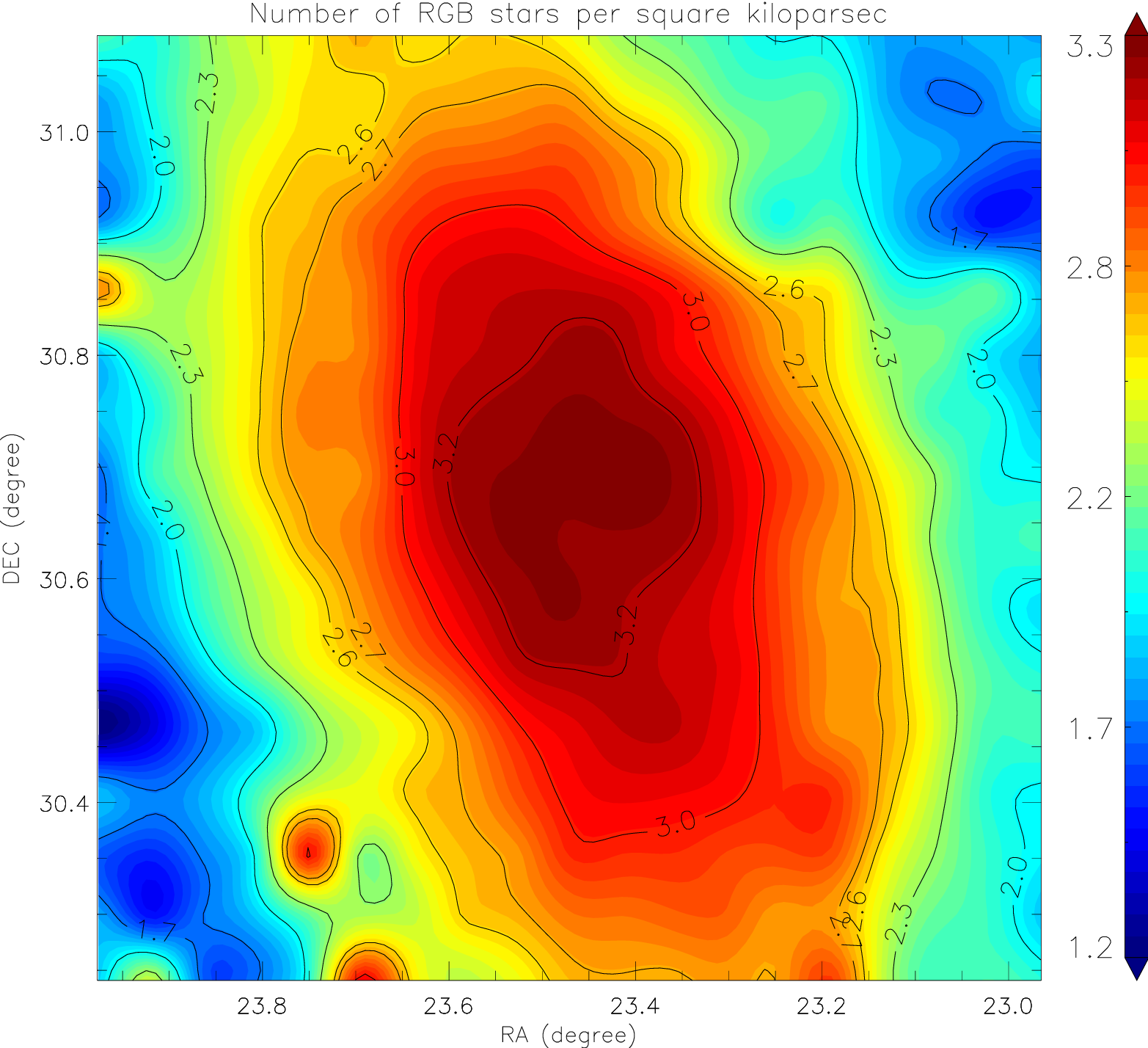,width=85mm}
\hspace*{3mm}
\psfig{figure=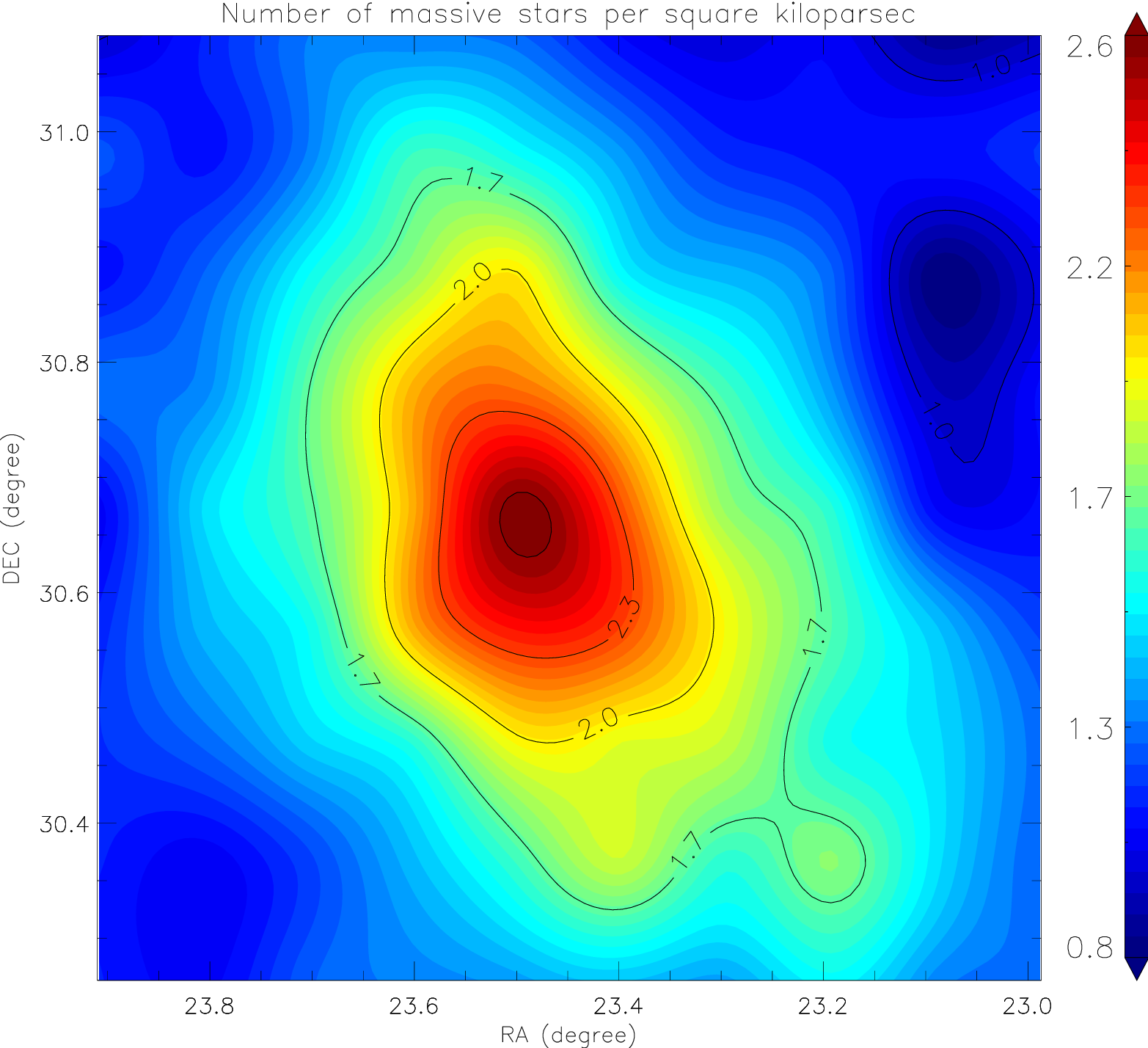,width=85mm}
}
\caption[]{Spatial distribution across M\,33 of the near-IR populations of
({\it Top left:}) WFCAM LPVs; ({\it Top right:}) AGB stars; ({\it Bottom
Left:}) RGB stars; ({\it Bottom right:}) massive stars. The units are
logarithmic in number of stars per square kpc.}
\end{figure*}

Maps of the surface density of the number of large-amplitude variable stars,
AGB stars, RGB stars and massive stars are presented in figure 24. The same
selection criteria as in Paper II are used to select these different
populations: the demarcation between massive stars and less-massive giant
stars is defined to run from $(J-K_{\rm s},K_{\rm s})=(0.6,18)$ mag to
$(J-K_{\rm s},K_{\rm s})=(0.9,15.6)$ mag, such that massive stars have colours
bluer than this (down to $K_{\rm s}=19.5$ mag) or have $K_{\rm s}<15.6$ mag,
whilst AGB stars and RGB stars are redder than this and have $16<K_{\rm s}<18$
mag or $18.3<K_{\rm s}<19.5$ mag, respectively.

The variable stars, AGB stars and massive stars are concentrated towards the
centre but the RGB stars do not show such strong central concentration. This
is in good agreement with what we found for the central square kpc of M\,33 in
Paper II. The distribution of the variable stars mostly mimics that of the AGB
stars, as expected, though the somewhat stronger central concentration in the
former suggests that more massive stars (AGB stars as well as RSGs) make a
larger contribution in the central part of M\,33 than further out in the disc.
Only hints can be seen of the spiral arm pattern in these maps.

\subsection{Cross-identifications in other catalogues}

We have cross-correlated our UKIRT/WFCAM variability search results with those
from two intensive optical monitoring campaigns (CFHT, Hartman et al.\ 2006,
in 5.3.1; DIRECT, Macri et al.\ 2001, in 5.3.2). We have also compared with
the optical catalogue of Rowe et al.\ (2005) which includes narrow-band
filters that they used to identify carbon stars (in 5.3.3); RSGs from Drout et
al.\ (2012) in 5.3.4; and a small number of Luminous Blue Variable (LBV) stars
from Humphreys et al.\ (2014) in 5.3.5. In addition, we compared our data with
the mid-IR variability search performed with the {\it Spitzer} Space Telescope
(McQuinn et al.\ 2007) in 5.3.6. The matches were obtained by search
iterations using growing search radii, in steps of $0\rlap{.}^{\prime\prime}1$
out to $1^{\prime\prime}$, on a first-encountered first-associated basis after
ordering the principal photometry in order of diminishing brightness
(K$_{\rm s}$-band for the UKIRT catalogue, i-band/I-band for the optical
catalogues, and 3.6-$\mu$m band for the {\it Spitzer} catalogue). Finally, we
examined our data on the recently identified 24-$\mu$m variables (Montiel et
al.\ 2014) which includes the giant H\,{\sc ii} region NGC\,604 (in 5.3.7).

\subsubsection{CFHT optical variability survey}

A variability survey of M\,33 was carried out with the 3.6-m
Canada--France--Hawai'i Telescope (CFHT) on 27 nights comprising 36 individual
measurements, between August 2003 and January 2005 (Hartman et al.\ 2006). Out
of two million point sources in a square-degree field, they identified more
than 1300 candidate variable blue and red supergiants, more than 2000
Cepheids, and more than 19\,000 AGB and RGB LPVs. Their catalogue comprises
Sloan g$^\prime$-, r$^{\prime}$- and i$^\prime$-band photometry to a depth of
$i^\prime\approx24$ mag.

\begin{figure}
\centerline{\psfig{figure=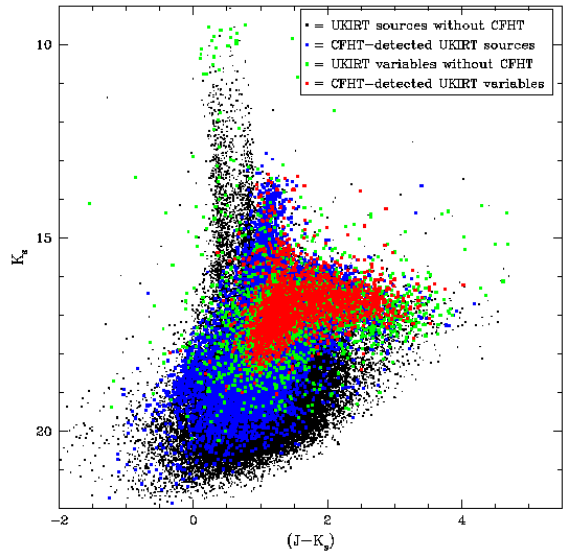,width=84mm}}
\caption[]{Near-IR CMD showing the stars from the WFCAM survey that were and
were not identified as variable stars in the CFHT optical variability survey
(Hartman et al.\ 2006).}
\end{figure}

Out of 36\,000 variable stars detected with CFHT in our field of M\,33, our
UKIRT/WFCAM survey detected 29\,600 stars (82\%), of which 1818 were found by
us to be variable (Fig.\ 25). Most of the CFHT variables that were missed in
our survey are fainter than the RGB tip; these are not the type of variables
that our survey aims to detect. The AGB and RSG variables that were not
detected in our survey have modest amplitudes (especially at IR wavelengths).
On the other hand, our survey detected some of the dustiest AGB variables that
were missed by the CFHT survey.

\subsubsection{DIRECT optical variability survey}

The DIRECT variability survey (Macri et al.\ 2001) aimed to determine a
distance estimate of M\,33 (and M\,31) using detached eclipsing binaries and
Cepheids. It was carried out on 95 nights with the F.L.\ Whipple observatory's
1.2-m telescope and on 36 nights with the Michigan--Dartmouth--MIT 1.3-m
telescope, between September 1996 and October 1997. Their catalogue contains
Johnson B- and V-, and Cousins I-band photometry for all stars with
$14.4<V<23.6$ mag, and the V-band $J$ variability index (cf.\ Section 3).

\begin{figure}
\centerline{\psfig{figure=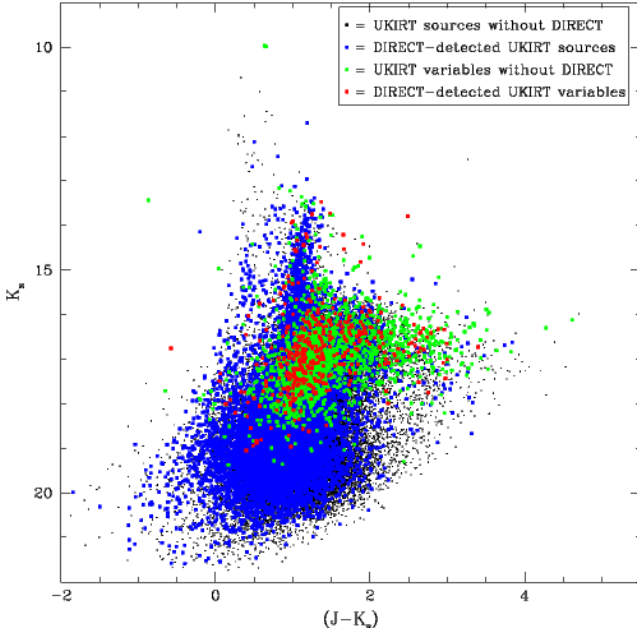,width=84mm}}
\caption[]{As Fig.\ 25, for the DIRECT optical variability survey (Macri et
al.\ 2001).}
\end{figure}

The central region of M\,33 observed in DIRECT is completely covered in our
WFCAM survey. The DIRECT survey listed 57\,581 stars among which 1383 have
$J>0.75$ (which the DIRECT team assumed as the variability threshold). Within
the coverage of DIRECT are located 116\,606 stars from the WFCAM catalogue,
including 1444 WFCAM variables. Of the 24\,422 stars from DIRECT that were
detected in our WFCAM survey, 412 were found by us to be variable. Hence, most
of the WFCAM variables were missed by the DIRECT survey (Fig.\ 26). This is
unsurprising since dusty AGB variables are faint at optical wavelengths, but
the relatively short duration of the DIRECT survey (little more than a year)
may have contributed to it missing also less dusty AGB variables.

\subsubsection{Carbon star survey}

Rowe et al.\ (2005) used the CFHT and a four-filter system in 1999 and 2000 to
cover much of M\,33. The filter system was designed to identify carbon stars
on the basis of their cyanide (CN) absorption as opposed to other red giants
that display titanium-oxide (TiO) absorption, using narrow-band filters
centered on 8120 and 7777 \AA, respectively. They added broad-band Mould V-
and I-band filters to aid in selecting cool stars. Carbon stars in their
scheme have [CN]$-$[TiO] $>0.3$ and $V-I>1.8$ mag, and M-type stars have
[CN]$-$[TiO] $<-0.2$ mag at the same V--I criterion (they only considered
stars for this purpose that had errors on these colours of $<0.05$ mag).

Within the area in common with our WFCAM survey, Rowe et al.\ detected
2\,079\,334 stars, from which 306\,292 stars (3.5\%) were identified with
WFCAM (a further 85\,098 stars from the WFCAM catalogue are outside of their
coverage). Among these, 11\,040 stars are M-type and 3134 are carbon stars. To
improve the cross-correlation and to avoid co-incidences because of the high
density of optical sources, a pre-selection was made on carbon stars and
M-type stars that are likely to have been detected at near-IR wavelengths. The
average offset between the two catalogues was found to be only
$0\rlap{.} ^{\prime \prime}02$ in both RA and DEC.

\begin{figure}
\centerline{\psfig{figure=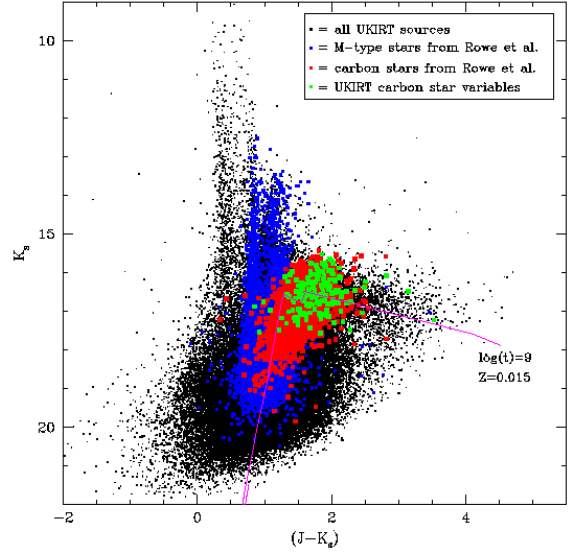,width=84mm}}
\caption[]{Near-IR CMD showing the M-type and carbon stars from the Rowe et
al.\ (2005) survey that were detected in our UKIRT/WFCAM survey. A 1-Gyr
isochrone from Marigo et al.\ (2008), for solar metallicity, is shown for
comparison.}
\end{figure}

The main giant branches -- viz.\ the RSGs around $(J-K_{\rm s})\approx0.8$ mag
and the AGB around $(J-K_{\rm s})\approx1$ mag -- are traced by M-type stars,
with carbon stars digressing towards redder colours at $K_{\rm s}\sim16$--17
mag (Fig.\ 27). The carbon star distribution is consistent with a typical
$t\sim1$ Gyr ($\log t\sim9$) isochrone or a little younger, i.e.\ birth masses
around 2--3 M$_\odot$. Some of the bright carbon stars show signs of reddening
to $(J-K_{\rm s})>2$ mag presumably due to circumstellar dust; as these are
among the brightest detected carbon stars, they must indicate the termination
point in carbon star evolution. Carbon stars are found fainter than the RGB
tip, down to $K_{\rm s}\sim19$ mag; these may have formed through binary mass
transfer. Such faint carbon stars are known in other, especially metal poor
populations, e.g., in the Sagittarius dwarf irregular galaxy (Gullieuszik et
al.\ 2007) or in the Galactic globular cluster $\omega$\,Centauri (van Loon et
al.\ 2007). Note that our WFCAM survey detected large-amplitude variability in
carbon stars only brighter than $K_{\rm s}\approx17.5$ mag, i.e.\ above the RGB
tip and consistent with thermal pulsing AGB stars that have become carbon
stars as a result of third dredge-up.

\begin{figure}
\centerline{\psfig{figure=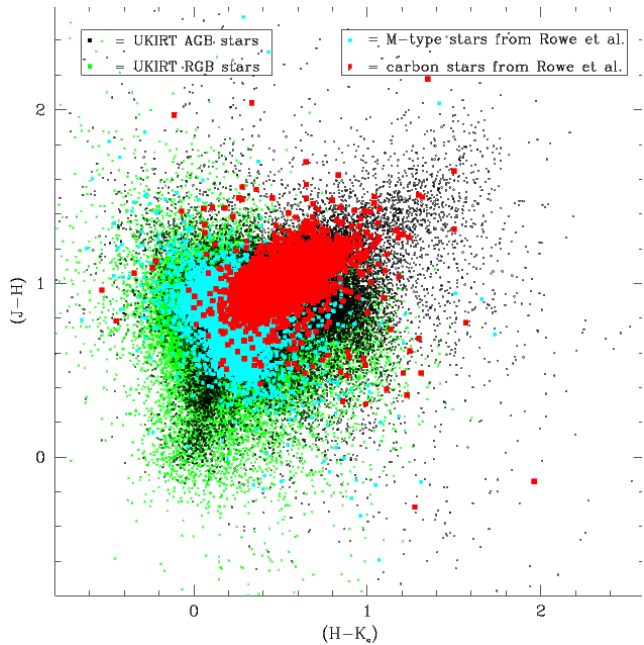,width=84mm}}
\caption[]{Near-IR colour--colour diagram showing the M-type and carbon stars
from the Rowe et al.\ (2005) survey that were detected in our UKIRT/WFCAM
survey. Stars with $18.5<K_{\rm s}<19.5$ mag and errors on the colours $<0.25$
mag are labeled as RGB stars, whilst stars with $16<K_{\rm s}<18$ mag and
errors on the colours $<0.25$ mag are labeled as AGB stars.}
\end{figure}

Figure 28 presents a near-IR colour--colour diagram for AGB and RGB stars
only, as other populations tend to confuse the picture in such diagram. Stars
with $18.5<K_{\rm s}<19.5$ mag are assumed to be on the RGB, whilst stars with
$16<K_{\rm s}<18$ mag are considered to be on the AGB (note that this selection
excludes some of either type but maximises the purity of these two samples).
For both selections we only kept stars with photometric errors on the colour
$<0.25$ mag. The M-type and carbon stars classified by Rowe et al.\ are
highlighted. Some of the M-type, but especially carbon stars follow part of a
sequence towards red colours, but the reddest of these, at $(H-K_{\rm s})>1$
mag and $(J-H)>1$ mag have generally not been bright enough at optical
wavelengths for spectral typing. These are dusty LPVs identified in our WFCAM
survey. The realm of the RGB stars spreads out over colour but generally this
happens in one but not both colours at once, suggesting blends and/or other
photometric uncertainties are to be blamed.

\subsubsection{Red Supergiant stars}

Drout et al.\ (2012) identified RSGs (and yellow supergiants) in M\,33 using
the Hectospec multi-fiber spectrograph on the 6.5-m Multiple Mirror Telescope
in two observing campaigns, in 2009 and 2010.

They divided their list in three different categories: those supergiants that
were selected both photometrically and kinematically, assigned rank 1; those
confirmed with just one method, assigned rank 2; and those which were not
selected by either method and which are likely foreground dwarfs, assigned
rank 3. They identified 189 rank-1 stars, 12 rank-2 stars and 207 rank-3
stars.

\begin{figure}
\centerline{\psfig{figure=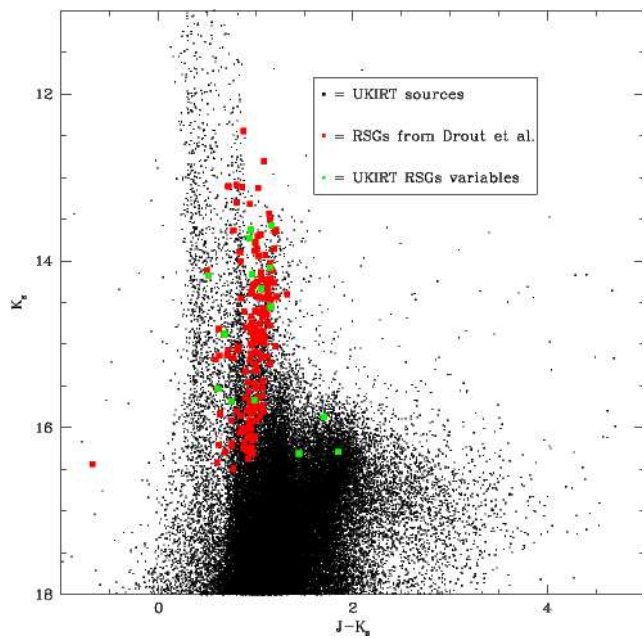,width=84mm}}
\caption[]{As Fig.\ 26, for the RSGs from Drout et al.\ (2012).}
\end{figure}

Our WFCAM survey detected 381 of the red stars in the survey by Drout et al.\
(93\%), from which 186 rank-1 stars (98\%) and 13 rank-2 stars. Of the rank-1
stars, 14 were found by us to be variable, as was one rank-2 star. Their RSGs
clearly delineate a branch in the near-IR CMD (Fig.\ 29; see also Fig.\ 30).
It is possible that some of the reddened sources are also RSGs but they will
have been too faint for spectral typing and hence not been included in the
work by Drout et al.\ (2012). Three of the variables in the top of the AGB
branch may instead be massive AGB stars or super-AGB stars (Fig.\ 29).

\subsubsection{The Luminous Blue Variable Var C}

Humphreys et al.\ (2014) identified a small number of LBVs in M\,33, one of
which is called ``Var C''. Since its discovery, a series of eruptions have
been witnessed in this star: 1940--1953 (Hubble \& Sandage 1953); 1964--1970
(Rosino \& Bianchini 1973); and 1982--1993 (Humphreys et al.\ 1988; Szeifert
et al.\ 1996). By 1998 it had returned to a minimum state (Burggraf et al.\
2014), but soon another, shorter episode of maximum light was seen from 2001
until 2005 (Viotti et al.\ 2006; Clark et al.\ 2012). Currently it is in a hot
quiescent stage (Humphreys et al.\ 2014).

\begin{figure}
\centerline{\psfig{figure=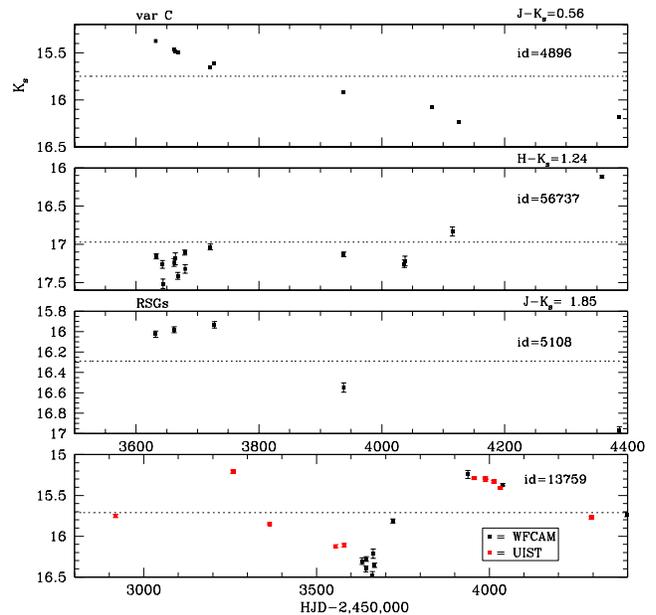,width=84mm}}
\caption[]{Light curves of selected variable stars: ({\it Top:}) the LBV Var
C; ({\it Middle:}) two of the RSGs; ({\it Bottom:}) one of the stars for which
we showed the lightcurve in Paper I, but now based on the combined UIST+WFCAM
photometry.}
\end{figure}

Var C was detected in our WFCAM survey four, five and ten times in the J-, H-
and K-band, respectively. The photometry is reliable since there is no
neighbouring star bright enough to seriously affect the photometry. The mean
magnitudes are $\langle J\rangle=16.32$ mag, $\langle H1\rangle=6.14$ mag and
$\langle K_{\rm s}\rangle=15.75$ mag; with $\langle J-K_{\rm s}\rangle=0.57$
mag it is one of the bluer confirmed variable stars in our list. Over the two
years of our monitoring campaign with WFCAM, Var C has steadily diminished its
brightness, though in 2007 it seems to have stabilised (Fig.\ 30).

\subsubsection{{\it Spitzer} mid-IR variability survey}

Five epochs of {\it Spitzer} Space Telescope imagery in the 3.6-, 4.5- and
8-$\mu$m bands have been analysed by McQuinn et al.\ (2007), to identify
variable stars using a smilar method to that we used ourselves.

\begin{figure}
\centerline{\psfig{figure=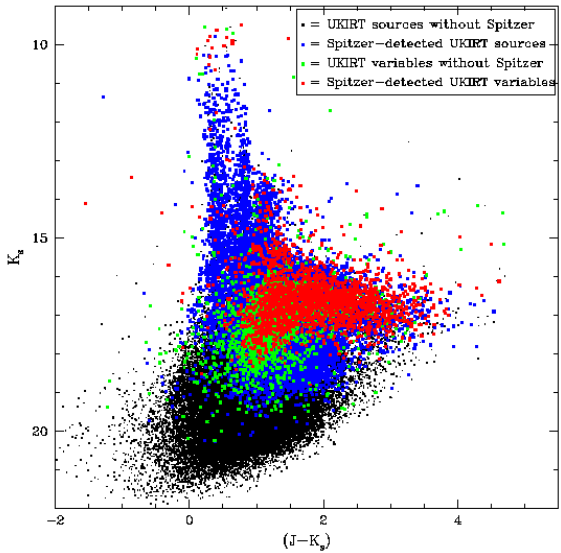,width=84mm}}
\caption[]{As Fig.\ 25, for the {\it Spitzer} mid-IR variability survey
(McQuinn et al.\ 2007).}
\end{figure}

The {\it Spitzer} images covered nearly a square degree, slightly larger than
our WFCAM survey; out of 40\,571 {\it Spitzer} sources, 2868 stars fall
outside the WFCAM monitoring coverage. Among the stars that {\it Spitzer}
detected within the region in common with our survey, 36\,411 are also in our
photometric catalogue, down to a little below the RGB tip (Fig.\ 31). Hence,
the recovery rate is $\sim97$\%. The recovery rate of the {\it Spitzer} survey
for RSGs, bright AGB stars and dusty AGB star variables from our WFCAM survey
is good. Blending is only problematic for stars within the central square kpc
of M\,33 (cf.\ Paper I), where the recovery rate decreases to $\sim70$\%.

\begin{figure}
\centerline{\psfig{figure=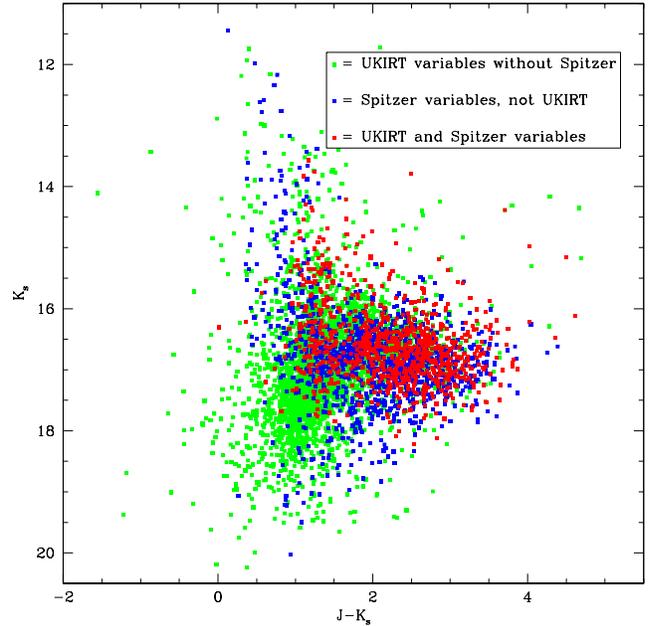,width=84mm}}
\caption[]{The variable stars that were or were not identified with WFCAM or
{\it Spitzer}.}
\end{figure}

Among the 2923 variable stars identified with {\it Spitzer}, 985 were
identified as variable stars also in our WFCAM survey. This corresponds to a
recovery rate of 35\% (excluding 113 stars that fall outside the WFCAM
coverage), i.e.\ slightly higher than the recovery rate by the WFCAM survey of
UIST variable stars in the central square kpc (cf.\ Section 3.1). On the other
hand, 3661 of our WFCAM variable stars had not been identified as variable
stars in the {\it Spitzer} survey. Both surveys do well in detecting variable
dusty AGB stars; the WFCAM survey is also sensitive to fainter, less dusty
variable red giants (Fig.\ 32) that were too faint for {\it Spitzer}.

\begin{figure}
\centerline{\psfig{figure=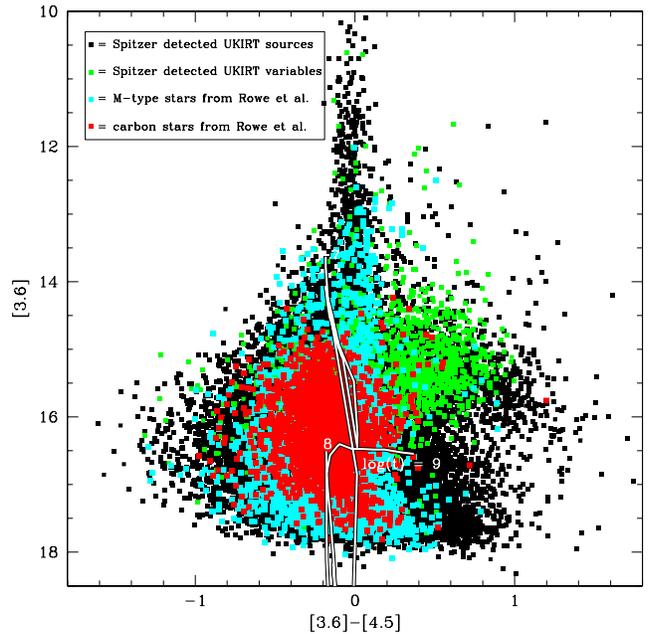,width=84mm}}
\caption[]{Mid-IR CMD of {\it Spitzer} photometry of our UKIRT/ WFCAM sources,
with WFCAM variables highlighted in green and M-type and carbon stars from
Rowe et al.\ (2005) in blue and red, respectively. Isochrones from Marigo et
al.\ (2008) for 10 Myr, 100 Myr and 1 Gyr are overlain for comparison.}
\end{figure}

Figure 33 shows a mid-IR CMD, with a well-populated sequence off from which a
branch extends towards redder colours and fainter 3.6-$\mu$m brightness
starting around $[3.6]\sim14$--16 mag and $([3.6]-[4.5])>0.2$ mag. This branch
mostly comprises dust-enshrouded objects, with a high fraction of WFCAM
variable stars. The WFCAM variables are plentiful also among the brighter
3.6-$\mu$m sources in the diagram; these are massive AGB stars and RSGs. The
M-type stars form a sequence of increasing 3.6-$\mu$m brightness, with carbon
stars located along part of this sequence. The near absence of M-type or
carbon stars in the red branch of dust-enshrouded objects is due to the
limited sensitivity of the optical spectral typing survey. The Padova
isochrones seem to underestimate the 3.6-$\mu$m brightness by a factor three
or so, most notable in the missing of the 1-Gyr isochrone of the red branch.

\subsubsection{Sources variable at 24 $\mu$m}

Montiel et al.\ (2014) have identified 24 sources in M\,33 that are variable
at a wavelength of 24 $\mu$m, based on {\it Spitzer} data with the MIPS
instrument. One of these is NGC\,604 (see below); the other sources could be
dusty evolved stars.

\begin{figure}
\centerline{\psfig{figure=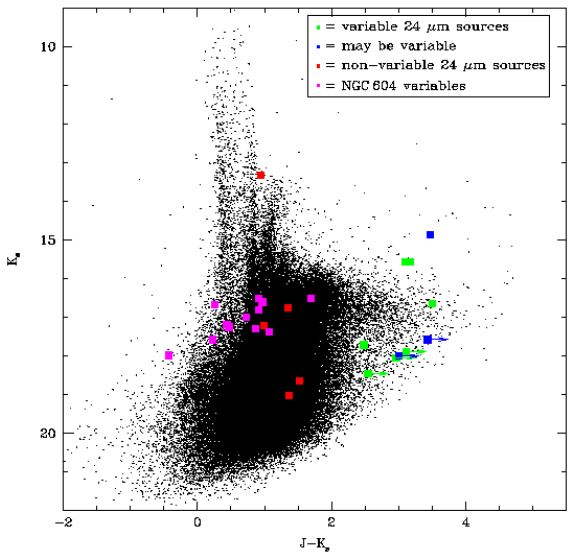,width=84mm}}
\caption[]{Near-IR CMD showing the 24-$\mu$m variables from Montiel et al.\
(2014) that were detected in our UKIRT/WFCAM survey. The variability
indication refers to our near-IR survey.}
\end{figure}

We find that VC6, 8, 9, 13, 16 21 and 23 (their nomenclature) are variable
with high confidence and VC10, 17 and 20 probably (but with fewer near-IR
epochs so less reliable). These are all red, with $(J-K_{\rm s})>2.4$ mag
(Fig.\ 34). We do not detect variability in VC14 but its photometry --
$(J-K_{\rm s})=0.94$ mag, $K_{\rm s}=13.32$ mag -- is consistent with being a
RSG and thus confirms the identification by Montiel et al.\ with the M2-type
760-d semiregular variable star VHK\,71 (van den Bergh, Herbst \& Kowal 1975);
VC17 is likely variable and its photometry -- $(J-K_{\rm s})=3.5$ mag,
$K_{\rm s}=14.87$ mag --  is consistent wiht it being a dusty RSG.

On the other hand, for VC7 we only find a moderately red star without detected
variability; this is probably the counterpart of the optical variable star
with a 143-d period with which Montiel et al.\ associated VC7, but not of the
24-$\mu$m variable itself as the 10-$\mu$m absorption in the latter appears
incompatible with such rapid variability.

None of the 24-$\mu$m variables with $([3.6]-[8.0])>4$ mag -- which includes
all of their far-IR detections -- were identified by us as near-IR variables.
If this is because they are in fact not dusty evolved stars then this would
explain why they do not obey the evolved stars sequences in their [3.6]--[8.0]
{\it vs.} [3.6]--[4.5] diagram; they may instead be young stellar objects.

\begin{figure}
\centerline{\psfig{figure=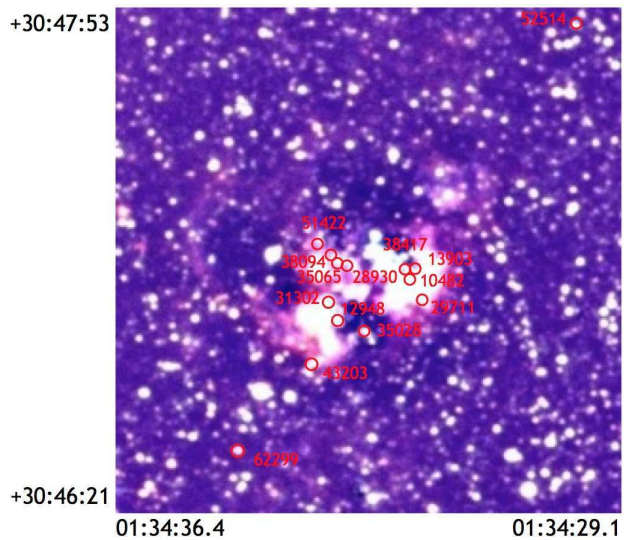,width=84mm}}
\caption[]{WFCAM JHK$_{\rm s}$ composite of the NGC\,604 H\,{\sc ii} region in
M\,33. The WFCAM variable stars are identified in red.}
\end{figure}

VC1 in Montiel et al.\ (2014) corresponds to NGC\,604, the second largest
extra-galactic H\,{\sc ii} region in the Local Group (after 30\,Doradus in the
LMC). It is a young star forming region with an age of 3--5 Myr (e.g., Wilson
\& Matthews 1995; Pellerin 2006). Located some 12$^{\prime}$ from the centre of
M\,33, it spans around 4.1 pc with a core--halo structure (Melnick 1980).
NGC\,604 has been studied throughout the electromagnetic spectrum, including
radio (Churchwell \& Grass 1999; Tosaki et al.\ 2007), infrared (Higdon et
al.\ 2003), optical (Tenorio-Tagle et al.\ 2004), ultraviolet (Keel et al.\
2004), and X-ray (Ma\'{\i}z-Apell\'aniz et al.\ 2004). These studies were
mostly concerned with the effect of the massive star formation on the
surrounding ISM. Using WFCAM we have identified 12 near-IR variable stars
within NGC\,604, which are shown in figure 35.

In the near-IR CMD (Fig.\ 34), the sources within NGC\,604 are neither
particularly luminous nor red -- in fact, most have $(J-K_{\rm s})<1$ mag. We
suspect that this group of variables may include any or all of the following
types of sources: (i) young early-type stars; (ii) stars whose photometry is
affected by nearby stars and/or nebulosity; and (iii) a dusty carbon star not
born in NGC\,604 (at $(J-K_{\rm s})=1.7$ mag).

\section{Conclusions}

WFCAM on UKIRT was used to extend our near-IR monitoring survey of the Local
Group spiral galaxy M\,33, from the central square kpc to a square degree.
K-band observations were complemented with occasional J- and H-band
observations to provide colour information.

The photometric catalogue comprises 403\,734 stars, among which 4643 stars
display large-amplitude variability. Investigation of the lightcurves and
location on the near-IR CMD with respect to theoretical models of stellar
evolution indicate that most of these stars are AGB stars or RSGs. They are
concentrated towards the centre of M\,33, more so than RGB stars. The majority
of very dusty stars which are heavily reddened even at IR wavelengths are
variable.

Cross-matching with optical monitoring campaigns and mid-IR variability
searches conducted with {\it Spitzer}, shows that the UKIRT/WFCAM catalogue of
variable stars is vastly more complete for the dusty variables than the
optical surveys, and more complete for less dusty variables than the {\it
Spitzer} surveys. Our catalogue is made publicly available at the CDS.

This work forms the basis for the next two papers in this series, to derive
the SFH (Paper V) and to measure the rate and location of dust production
(Paper VI) across M\,33. 

\section*{Acknowledgments}

We thank the staff at UKIRT for their excellent support of this programme, and
the referee for her/his constructive report. JvL thanks the School of
Astronomy at IPM, Tehran, for their hospitality during his visits. We are
grateful for financial support by The Leverhulme Trust under grant No.\
RF/4/RFG/2007/0297, by the Royal Astronomical Society, and by the Royal
Society under grant No.\ IE130487.


\label{lastpage}

\end{document}